\documentclass[12pt, draftclsnofoot, onecolumn]{IEEEtran}

\usepackage{cite}

\ifCLASSINFOpdf
\usepackage[pdftex]{graphicx}
\else
\fi
\usepackage{amsmath}
\usepackage{array}
\usepackage{fixltx2e}

\usepackage{stfloats}
\usepackage{epsfig}
\usepackage{amssymb, amsmath}
\usepackage{color}
\usepackage{hyperref}
\usepackage{graphicx}
\usepackage{color}
\usepackage{graphicx}
\usepackage{tabu}

\usepackage[subrefformat=parens,labelformat=parens]{subfig}
\usepackage{float}
\usepackage{amssymb}
\usepackage{amsmath,amsfonts,amssymb}
\usepackage{leftidx}
\usepackage{verbatim}
\usepackage{stfloats}
\usepackage{comment}

\newtheorem{theorem}{Theorem}

\newtheorem{lemma}{Lemma}

\newtheorem{example}{Example}
\begin{document}
%

\title{Designing MRC Receivers to Manage Unknown Timing Mismatch-A Large Scale Analysis}
%
%
%

\author{Mehdi~Ganji,~\IEEEmembership{Student Member,~IEEE,}
Hamid~Jafarkhani,~\IEEEmembership{Fellow,~IEEE} \thanks{M. Ganji and H. Jafarkhani are with the Center for Pervasive Communications and Computing, University of California, Irvine, CA, 92697 USA (e-mail:
\{mganji, hamidj\}@uci.edu). This work was supported in part by the NSF Award CCF-1526780.}}

\maketitle

\begin{abstract}
There has been extensive research on the large scale multiuser multiple-input multiple-output (MU-MIMO) systems recently. However, there are many obstacles in the way to achieve full potential of using large number of receive antennas. One of the main issues, which will be investigated thoroughly in this paper, is timing asynchrony among signals of different users. Most of the works in the literature assume that the received signals are perfectly aligned, which is not practical. We quantify the uplink achievable rates obtained by the MRC receiver with perfect channel state information (CSI) and imperfect CSI while the system is impaired by unknown time delays among received signals. Then, we use these results to design new algorithms in order to alleviate the effects of timing mismatch. We also analyze the performance of the introduced receiver design, which is called MRC-ZF. To verify our analytical results, we present extensive simulation results which thoroughly investigate the performance of the traditional MRC receiver and the introduced MRC-ZF receiver. \\
\end{abstract}


%

\section{Introduction}
%
%
%
%
Introducing multiple-input multiple-output (MIMO) systems was a breakthrough in communication systems. MIMO communications was studied extensively during the past two decades \cite{bolcskei2006space,duman2008coding,jafarkhani2005space}. Using multiple antennas at the transmitter and the receiver provides the opportunity to increase the capacity and improve the performance significantly \cite{foschini1996layered, tarokh1999space}. One of the applications of MIMO systems is in multiuser scenarios where $K$ users, each equipped with multiple antennas, communicate with a common multiple antenna receiver. Besides traditional problems in point to point communication, due to the distributed nature of multiuser-MIMO (MU-MIMO) systems, new challenges exist like timing mismatch between received signals from different users \cite{verdu1998multiuser}. When the number of users and the number of receive antennas are moderate, this issue is often handled by synchronization methods\cite{van1999time,zhou2005synchronization, timing_recovery}. Recently, it has been shown that timing mismatch can even improve the performance when the timing mismatch values are known by the receiver and proper sampling and detection methods are used \cite{das2011mimo,lu2012asynchronous,poorkasmaei2015asynchronous,avendi2015differential,mehdi}. However, increasing the number of receive antennas and users makes the time delay estimation or synchronization challenging, especially in the context of massive MIMO systems\cite{massiveMIMO}.

In large scale MU-MIMO systems, the base station is equipped with very large number of receive antennas and communicates with tens of users at the same time and frequency. The benefits of the  massive MIMO settings including, near optimal performance using simple processing like maximum ratio combining (MRC), increased spectral efficiency and energy efficiency, have been studied in the literature\cite{larsson2014massive,ngo2013energy,lu2014overview}. However, there are many challenges which need to be addressed before the gains can be realized in practice\cite{rusek2013scaling}, \cite{rogalin2014scalable}. For hundreds of receive antennas, one major challenge is the fact that it is impossible to receive perfectly aligned signals at all the receive antennas, especially in a distributed scenario where the receive antennas are not collocated. In \cite{motivation}, the authors showed that when multiple base stations (BSs) communicate with their corresponding users, the interference is inherently asynchronous, meaning that BSs cannot align all the interfering signals at each
user because of the different propagation times between the
BSs and users. This is exactly what happens in uplink, where multiple users communicate with base station with multiple distributed receive antennas. Even if the users perform timing correction using the timing advance (TA) estimate received in the physical downlink control channel (PDCCH), the synchronization can be realized at only one receive antenna. However, due to different propagation delays, the other receive antennas will experience asynchrony. Therefore, it is of great importance to investigate timing mismatch in large scale MU-MIMO systems.

\begin{figure*}[!t]
\centering
\subfloat[ISI caused by multi-path channel]{\includegraphics[width=2.5in]{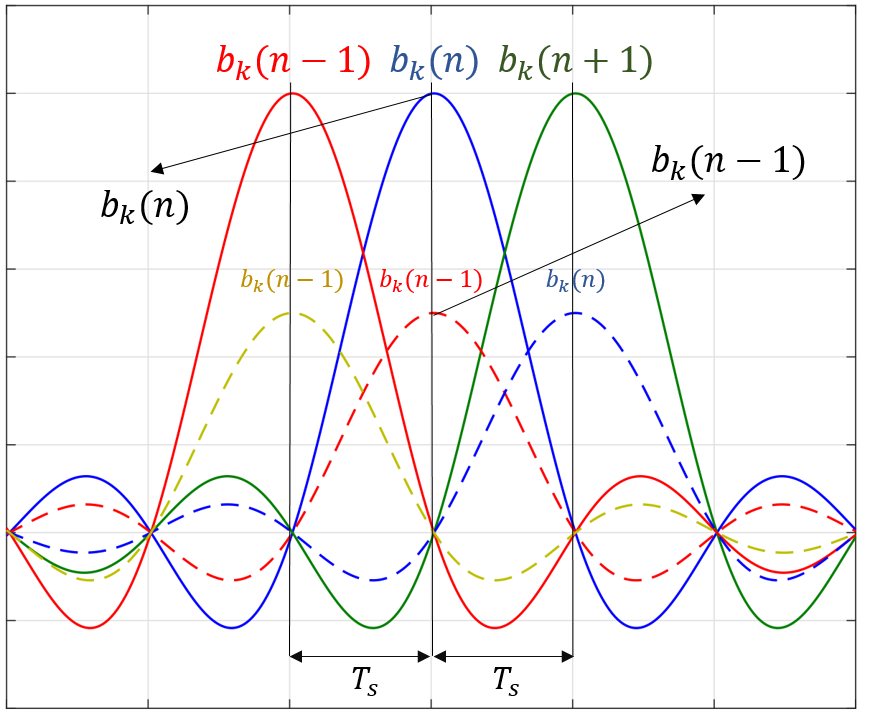}
\label{sampling_multipath}}
\hfil
\subfloat[ISI caused by imperfect sampling]{\includegraphics[width=2.5in]{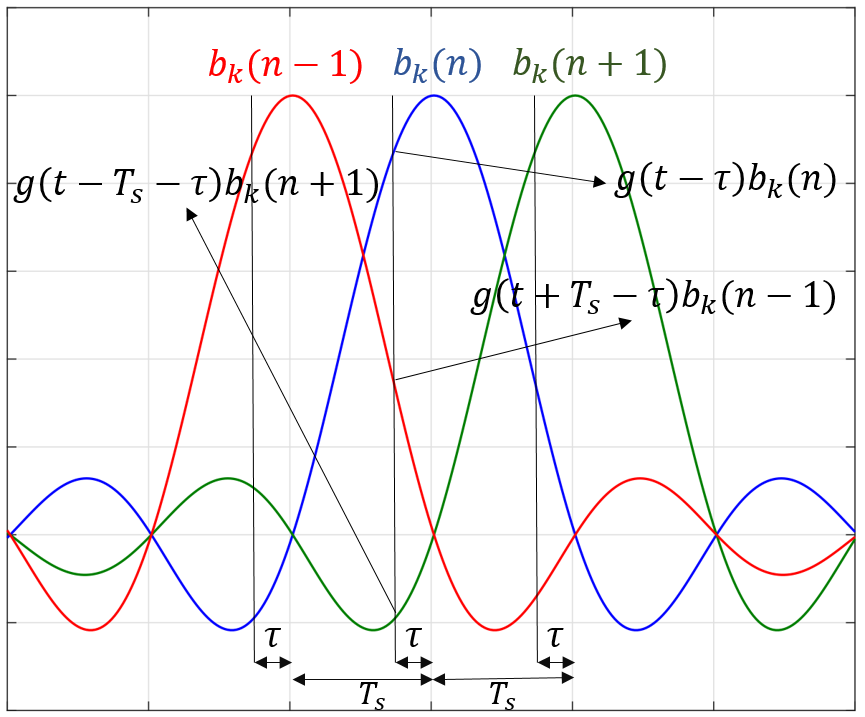}
\label{sampling_mismatch}}
\caption{Demonstration of two different sources of creating ISI}
\label{fig_sim}
\end{figure*}

For large scale multi-carrier MU-MIMO systems, the timing mismatch between the received signals can be modeled as the phase rotation of the received symbols. Such a phase rotation behaves similar to the phase noise introduced by the oscillator at the receiver and has been studied in the literature \cite{pitarokoilis2012effect}, \cite{ krishnan2016linear}. However, to the best of our knowledge, there is no work in the literature to consider the timing mismatch in single-carrier massive MIMO scenarios. In \cite{pitarokoilis2012optimality}, it is shown that using single-carrier transmission can achieve a near optimal sum rate. The authors have proposed a simple precoding which mitigates the inter-symbol interference (ISI) caused by channel multi-path. However, they assume perfect symbol-level alignment enabling perfect sampling at the peak point of the transmitted pulse shape which might be challenging in a large scale MU-MIMO system. Inevitable timing mismatch between received signals, results in imperfect sampling, and hence creates another source of ISI as illustrated in Fig. \ref{fig_sim}. 
In Fig. \subref*{sampling_multipath}, the ISI is created by dotted symbols which are delayed copies of the transmitted symbols caused by a frequency selective channel. In Fig. \subref*{sampling_mismatch}, the ISI is generated by imperfect sampling. If the timing mismatch values are known at the receiver, ISI-free samples can be obtained for each user by oversampling as many times as the number of users, as explained in \cite{mehdi}. However, considering practical challenges for delay acquisition in a large scale MU-MIMO system, we assume that the timing mismatch values are unknown and the receiver only knows their distribution.

It is shown in the literature that in large scale MU-MIMO systems, a low complexity MRC receiver can approach near optimal performance and even outperforms its complex counterparts, i.e., ZF and MMSE receivers, at low SNR\cite{ngo2013energy}. The MRC receiver also follows the power scaling law. The power scaling law roughly indicates that, for all SNRs, to maintain the same quality-of-service of a single-user
SISO scenario with no interference, the transmitting power of single-antenna users to a 100-antenna BS would be almost 1 \% of that of the single-user SISO system \cite{rate_massiveMIMO}. As we shall see, ignoring the asynchrony can significantly degrade the performance of the MRC receiver. We develop a mathematical model that explicitly accounts for the timing mismatch among the received signals. We then quantify the detrimental impact of asynchrony on the MRC receiver, and suggest how to mitigate it by making some modifications to the MRC receiver. We have the following specific contributions:
\begin{itemize}
\item We derive a tight approximation for the achievable uplink rate using the MRC receiver in the presence of timing mismatch. We consider a single cell scenario with perfect and imperfect channel state information (CSI). Our results are general and cover any arbitrary delay distribution including the synchronous scenario.
\item We find the optimal sampling times which maximize the asymptotic achievable rate by the MRC receiver when the number of receive antennas goes to infinity. 
\item We show that the MRC receiver cannot provide the power scaling law when there is misalignment between received signals.
\item We introduce a new receiver design called MRC-ZF which alleviates the effects of timing mismatch and follows the power scaling law. 
\item We derive an achievable uplink rate approximation when the MRC-ZF receiver is used. 
\end{itemize}

The rest of the paper is organized as follows: first we introduce the system model and explain discretization and receiver processes in Section \ref{sec-sys}. We analyze the achievable rates obtained by the MRC receiver when unknown delays exist in Section \ref{sec-rate} and then MRC-ZF receiver structure and corresponding achievable rates are presented in Section \ref{sec-MRCZF}. Next, simulation results are presented in Section \ref{sec-simu} to verify the effectiveness of our proposed methods. Finally, we summarize our contributions in Section \ref{sec-con}.\\

\section{System Model}\label{sec-sys}
\subsection{Received Signal Model}
We consider a system with $K$ single antenna users, transmitting data to a common receiver with $M$ receive antennas simultaneously, where $M$ can correspond to a massive deployment scenario. The signal transmitted from User $k$ is described by:
\begin{eqnarray}
&s_k(t)=\sqrt{\rho_d}\sum_{i=1}^{N}{b_k(i)p(t-(i-1)T_s)} \ \ 
\end{eqnarray}
where $T_s$, $\rho_d$ and $p(.)$ represent the symbol period, the transmit power, and the pulse-shaping filter with non-zero duration of $T$, respectively. For the rectangular pulse shape, $T=T_s$, and for Nyquist pulse shapes truncated with $I$ significant adjacent side lobes, $T=2(I+1)T_s$. The number of significant adjacent side lobes is specified based on the desired stop band attenuation. More number of side lobes results in higher stop band attenuation. For example, we assume 3 side lobes in our simulation which results in about $25$ dBm stop band attenuation for the square-root raised cosine pulse shape. Also, $N$ is the frame length and $b_k(i)$ is the transmitted symbol by User k in the $i$th time slot. The transmitted signals are received with a relative delay of $d_{km}T_s+\tau _{km}$ and a channel path gain of $c_{km}=\sqrt{\beta_k} h_{km}$, where $\sqrt{\beta_k}$ is the path-loss that depends on the distance between the corresponding user and the base station and $h_{km}$ represents the fading coefficients. We focus on the flat fading channel in this work and postpone the analysis of the frequency selective channel to our future work. We assume the fading coefficients are independent due to having sufficiently separated receive antennas and follow Rayleigh distribution with zero mean and variance one, however, the fading can be  modeled by more general scenarios assuming spatial correlation among the receive antennas which will be considered in our future work.
Then, the continuous received signal at the $m$th receive antenna can be represented by:
\begin{eqnarray}
\label{eq:con}
&\psi_{m}(t)=\sum_{k=1}^{K}{c_{km}s_k(t-d_{km}T_s-\tau _{km})}+\nu_{m}(t) 
\end{eqnarray}
where $K$ is the number of users and $\nu_m(t)$ is the white noise with zero mean and variance one.  In this work, we assume that the values of frame asynchrony, i.e., $d_{km}$ is known at the receiver \cite{timing_recovery}. Therefore, without loss of generality, we assume $d_{km}=0$ in Eq. (\ref{eq:con}). However, the symbol-level asynchrony, i.e., $\tau_{km}$ is unknown and is treated as a random variable between $[0, T_s]$. The reason behind this assumption is that the precision needed for the symbol-level synchronization is much higher than the frame synchronization, which becomes important particularly in massive MIMO context, where using estimation algorithms with high precision and thus long acquisition time is infeasible. Also, the value of $d_{km}$ is discrete and easily representable by a finite number of bits, while $\tau_{km}$ is continuous and its exact representation needs infinite number of feedback bits which is impossible. Therefore, always, there will be some residual error that can be modeled as an unknown random variable $\tau_{km}$. 

In general, we only need to know the joint distribution of time delays to calculate the achievable rates and design our proposed algorithms. Although the time delays can follow any joint distribution, we consider the i.i.d case with the following distribution:
\begin{flalign}\label{delays}
f(\tau)= \frac{1}{K}\delta(\tau)+\frac{K-1}{K}U(0,T_s)
\end{flalign}
Eq. (\ref{delays}) reflects the fact that our desired user is received with probability of $\frac{1}{K}$ as the first user and with probability of $\frac{K-1}{K}$ with a time delay, uniformly distributed on $[0, T_s]$, with respect to the first received user. Because of our probabilistic framework, the order of users is not important and the receiver only needs to know when the received signal starts, however, it does not need to know which user is the first one.

\subsection{Matched Filter's Output Signal Model}
In this section, we explain the receiver design that includes the transformation of the continuous signal in Eq. (\ref{eq:con}) into discrete samples and the combination of the obtained samples at different receive antennas by the MRC method. To obtain the discrete samples of the received signal, first, the continuous received signal should be passed through a matched filter and its output can be written as follows:
\begin{flalign}
\hat{\psi}_{m}(t)&=\sqrt{\rho_d}\sum_{k=1}^{K}{\sqrt{\beta_k} h_{km}\sum_{i=1}^{N}{b_k(i)g(t-(i-1)T_s-\tau_{km})}}+\nu_{m}(t) \ast p(t)
\end{flalign}
where $g(t)=p(t)\ast p(t)$. The convolution $g(t)$, called the convolved pulse shape, is zero outside the interval of $[0, 2T]$. Then, the output of the matched filter is sampled at the instants of $t^s_{n}$, which are equal to $eT_s+T+(n-1)T_s, n=1, \cdots,N$. The quantity of $e \in [0,1]$ is a design parameter that affects the performance, significantly. If all the received signals were synchronized, then $e=0$ would be the optimum value, which is the assumption in most of the work in the literature. However, due to having unknown delays among received signals, the optimum value of $e$ is not zero anymore and will be found based on the system model characteristics. The obtained samples at the sampler of the $m$th receive antenna, denoted by $y_{m}(n)=\hat{\psi}_{m}(t)|_{t^s_n}$, can be written as:
\begin{flalign}\label{samp_out}
y_{m}(n)=\sqrt{\rho_d}\sum_{k=1}^{K}{\sqrt{\beta_k} h_{km}\sum_{i=1}^{N}{b_k(i)g(eT_s+T+(n-i)T_s-\tau_{km})}}+{\hat \nu}_{m}(t)\big|_{eT_s+T+(n-1)T_s} 
\end{flalign}
where ${\hat \nu}_{m}(t)=\nu_{m}(t) \ast p(t)$. We can put the obtained samples together and form the system model equation as follows:
\begin{flalign}
\label{sys}
\boldsymbol{y_{m}}=\sqrt{\rho_d }\sum_{k=1}^{K}{\sqrt{\beta_k} h_{km}\boldsymbol{G_{km}}\boldsymbol{b_k}}+\boldsymbol{n_{m}} 
\end{flalign}
where $\boldsymbol{b_k}=[b_k(1),b_k(2),\cdots, b_k(N)]^T$ is the transmitted frame by the $k$th user and $\boldsymbol{ n_{m}}=[n_{m}(1),n_{m}(2),\cdots, n_{m}(N)]^T$ is the noise vector containing samples of $\hat \nu_{m}(t)$, i.e., $ n_{m}(n)=\hat \nu_{m}(t)|_{eT_s+T+(n-1)T_s}, 1 \leq n \leq N$. Also, $\boldsymbol{G_{km}}$ is an $N\times N$ matrix defined as:
\small
\begin {flalign}\label{gg}
\boldsymbol{G_{km}}=\begin{pmatrix}
g(eT_s+T-\tau_{km})&\cdots & g(eT_s+T+(1-N)T_s-\tau_{km})\\ 
g(eT_s+T+T_s-\tau_{km})& \cdots & g(eT_s+T+(2-N)T_s-\tau_{km}) \\ 
\vdots & \ddots & \vdots\\ 
g(eT_s+T+(N-1)T_s-\tau_{km})&\cdots &g(eT_s+T-\tau_{km})
\end{pmatrix}_{N\times N}
\end{flalign}
\normalsize
Defining $\boldsymbol{T_{km}}=\sqrt{\beta_k} h_{km}\boldsymbol{G_{km}}$, Eq. (\ref{sys}) can be written in the following short form:
\begin{flalign}
\boldsymbol{y_{m}}=\sqrt{\rho_d} \sum_{k=1}^{K}{\boldsymbol{T_{km}}\boldsymbol{b_k}}+\boldsymbol{n_{m}} 
\end{flalign}
The noise vector has zero mean and its covariance matrix is the identity matrix because the pulse shapes are normalized and satisfy the Nyquist ISI-free condition.

After obtaining the samples, they are combined using the well-known MRC method. Denoting $\tilde{c}_{lm}$ as the estimate of the channel coefficient between the $l$th user and the $m$th receive antenna, the MRC output for detection of the $l$th user's symbols, i.e., $\boldsymbol{y_l^{mrc}}=\frac{1}{M}\sum_{m=1}^M{\tilde{c}^*_{lm}\boldsymbol{y_{m}}}$, can be expressed as:
\begin{flalign}\label{sys_mrc}
\boldsymbol{y_l^{mrc}}=\sqrt{\rho_d} \sum_{k=1}^{K}{\boldsymbol{T^{mrc}_{lk}}\boldsymbol{b_k}}+\boldsymbol{n^{mrc}_l}
\end{flalign}
where the effective channel matrices and the resulting noise vector are denoted by $\boldsymbol{T^{mrc}_{lk}}$ and $\boldsymbol{n^{mrc}_{l}}$, respectively, and will be defined later based on the available CSI and detection methods. In the next section, we analyze the performance of the MRC detection with perfect CSI and estimated CSI at the receiver. 
\section{The Achievable Rate of MRC Receiver}\label{sec-rate}
\subsection{Perfect CSI}

Here, we assume that channel coefficients are estimated separately for each user. It might be impractical due to the large number of users being served by the base station; however, it will uncover the main effects of unknown time delays on the performance. By assuming $\tilde c_{km}=c_{km}$, the effective channel matrix, i.e., $\boldsymbol{T^{mrc}_{lk,p}}$ and effective noise vector, i.e., $\boldsymbol{n_{l,p}^{mrc}}$ can be represented as follows, respectively:\footnote{Throughout the paper, the subscripts $p$ and $ip$ are used for perfect CSI and imperfect CSI, respectively.} 
\small
\begin {flalign}\label{effective_matrix}
\boldsymbol{T^{mrc}_{lk,p}}=\frac{1}{M}\sum_{m=1}^{M}\sqrt{\beta_l \beta_k}h^*_{lm}h_{km}\boldsymbol{G_{km}},\ \ 
\boldsymbol{n_{l,p}^{mrc}}=\frac{\sqrt{\beta_l}}{M}\sum_{m=1}^M h^*_{lm} \boldsymbol{n_m}
\end{flalign}
\normalsize
The achievable rate for the corresponding system model is denoted in the next theorem.
\begin{theorem}\label{11}
The achievable rate of the MRC receiver for User $l$, when there is unknown time delays between received signals, can be approximated as:
\small
\begin{flalign}
\tilde R_{l,p}^{mrc}\approx\log_{2}{\left(1+\frac{\rho_d \beta_lE^2[g_0]}{\rho_d\sum\limits_{i=-I}^{I}{E[g_i^2]}\sum\limits_{\substack{k=1\\ k\neq l}}^{K}{\beta_k}+\rho_d \beta_l\sum\limits_{\substack{i=-I }}^{I}(2E[g_i^2]+(M\bar\delta[i]-1)E^2[g_i])+1}\right)}
\label{mrc_mapping}
\end{flalign}
\normalsize
where 
\begin{flalign}\label{eq:int}
{E}[g_i^{n}]=\int_{-\infty}^{\infty}{g^n(eT_s+T+iT_s-\tau)f(\tau)d\tau}
\end{flalign}
Also $\bar\delta[i]=1-\delta[i]=\left\{\begin{matrix}
1 \ \ i\neq 0\\ 
0 \ \ i=0
\end{matrix}\right.$, and $I$ is the number of significant adjacent side lobes of the pulse shape.
\end{theorem}
\begin{IEEEproof}
The proof is presented in Appendix \ref{appendix_1}.
\end{IEEEproof}
The first term in the denominator of Eq. (\ref{mrc_mapping}) is the inter-user interference (IUI) caused by other users. The second term is the inter-symbol interference (ISI), caused by the adjacent symbols of the desired user, and also the uncertainty in the coefficient of the desired symbol. The last term is related to the additive white noise.

\begin{example}\label{example_2}
In this example, we find the values of ${E}[g_i^{n}]$ in Eq. (\ref{eq:int}) for the rectangular pulse shape and the time delay distribution presented in Eq. (\ref{delays}):
\small
\begin{flalign}
\nonumber
\left\{\begin{matrix}
E[g_0]=\frac{1}{K}\left(1-{e}\right)+\frac{K-1}{K}\left(\frac{1}{2}+{e}-{e^2}\right)\\ 
E[g_{-1}]=\frac{1}{K}\left({e}\right)+\frac{K-1}{K}\left(\frac{e^2}{2}\right)\\ 
E[g_1]=\frac{K-1}{K}\left(\frac{(1-e)^2}{2}\right)
\end{matrix}\right\} ,\ \left\{\begin{matrix}
E[g^2_0]=\frac{1}{K}\left(1-{e}\right)^2+\frac{K-1}{K}\left(\frac{1}{3}+{e}-{e^2}\right)\\\ 
E[g^2_{-1}]=\frac{1}{K}\left({e^2}\right)+\frac{K-1}{K}\left(\frac{e^3}{3}\right)\\ 
E[g^2_1]=\frac{K-1}{K}\left(\frac{(1-e)^3}{3}\right)
\end{matrix}\right\} 
\end{flalign}
\normalsize
Note that for the rectangular pulse shape, $E[g_i]$ and $E[g_i^2]$ are nonzero only when $i=-1,0,1$. After calculating these values for different values of $e$ and $K$, they can be inserted into Eq. (\ref{mrc_mapping}) to find the corresponding achievable rates. 
\end{example}

In the ideal case of synchronized reception, the power scaling law of massive MIMO systems states that the transmit power of each user can be cut down by $\frac{1}{M}$ with no degradation in the achievable rate of each user, i.e., $R_{l,p}^{mrc-ideal}\rightarrow\log_{2}{(1+E_d\beta_l)}\ as \ M\rightarrow \infty, \rho_d=\frac{E_d}{M}$ \cite{ngo2013energy}. However, by ignoring the inevitable timing mismatch, the promised benefit of power scaling in a massive MIMO setting vanishes. In more details, if we put $\rho_d=\frac{E_d}{M}$ in Eq. (\ref{mrc_mapping}) and let $M$ go to infinity, then we will have:
\small
\begin{flalign}\label{bala}
R_{l,p}^{mrc}\rightarrow\log_{2}{\left(1+\frac{E_d \beta_l E[g_0]^2}{E_d \beta_l \sum\limits_{\substack{i=-I \\ i\neq 0}}^{I}{E[g_i]^2+1}}\right)}
\end{flalign}
\normalsize
The achievable rate in Eq. (\ref{bala}) is limited by ISI, and by increasing the transmit power it will be saturated to a constant value, i.e.:
\small
\begin{flalign}\label{pain}
R_{l,p}^{mrc}\rightarrow\log_{2}{\left(1+\frac{E[g_0]^2}{\sum\limits_{\substack{i=-I \\ i\neq 0}}^{I}{E[g_i]^2}}\right)}
\end{flalign}
\normalsize
Therefore, at high SNR regime, no matter how much transmit power is used, the achievable rate converges to a fixed value independent of the transmit power. This fixed value depends on the delay distribution, the pulse shape and $e$. Using this criterion, the value of $e$ can be optimized for any given pulse shape and time delay distribution. For example, the optimum value of $e$ for the rectangular pulse shape and the delay distribution presented in Eq. (\ref{delays}) can be found by optimizing the following expression:
\begin{flalign}
\max_{e}{\frac{\left(2(1-e)+(K-1)(1-2e^2+2e)\right)^2}{(2e+(K-1)e^2)^2+(K-1)^2(1-e)^4}}
\end{flalign}
which is obtained by inserting the values of $E[g_i]$, found in Example \ref{example_2}, into Eq. (\ref{pain}). The optimal values of $e$ for a few examples of $K$ are presented in Table \ref{table1}. 
\begin{table}[h!]
\caption{Optimal Sampling Origin $e$} 
\centering 
\begin{tabu}{c c c c c c c c c}
\hline\hline 
Case & $K=2$ & $K=4$ & $K=6$ & $K=8$ & $K=10$ & $K=12$ & $K=14$ & $K=16$ \\ 
\hline 
Optimal $e$ & 0.18 & 0.35 & 0.41& 0.44 & 0.45 & 0.46 & 0.46 & 0.47\\ 
\hline 
\end{tabu}
\label{table1} 
\end{table}
Note that since we have no knowledge of $\tau$s and they are assumed to follow a random distribution, optimum value of $e$  does not depend on each realization of time delays anymore. As the value of $K$ increases, the delay distribution approaches to a uniform distribution. For uniform distribution, the optimal value of $e$, which maximizes the ratio of the expectation of the signal power to the expectation of the interference plus noise power, is equal to half, i.e., the midpoint of the symbol interval. The simulation results for the root raised cosine pulse shape are presented in Section \ref{sec-simu}.

\subsection{Imperfect CSI}{\label{zohr}}
In this section, we assume the channel coefficients are estimated by sending known sequences of symbols, called pilot sequences. Each user assigns its first $N_p$ symbols of each frame to send pilot symbols. We denote the assigned pilot sequence to the $k$th user as $\boldsymbol{p_k}=[p_k(1),\cdots,p_k(N_p)]$. It is common in the literature that the assigned pilot sequences for different users are mutually orthogonal, i.e., $\langle\boldsymbol{p_i}.\boldsymbol{p_j}\rangle=\delta[i-j]$, where $\langle\ .\ \rangle$ shows the inner product. In addition, $N_p$ should be equal to or greater than the number of users and its optimal value is shown to be $N_p=K$ \cite{hassibi}. The mutual orthogonality enables all the users to send the pilot symbols simultaneously without interfering with each other. The $K \times N_p$ matrix that contains all the pilot sequences is represented by:
\small
\begin{flalign}
\boldsymbol{\Phi}=\begin{pmatrix}
p_1(1)&\cdots & p_1(N_p)\\ 
p_2(1)&\cdots & p_2(N_p) \\ 
\vdots & \ddots & \vdots\\ 
p_K(1)&\cdots & p_K(N_p)
\end{pmatrix}_{K\times N_p}
\end{flalign}
\normalsize
Due to orthogonality between rows, the pilot matrix is unitary, i.e., $\boldsymbol{\Phi}\boldsymbol{\Phi}^H=\boldsymbol{I_{K}}$. In the ideal case of perfect synchronization, the received signal can be written as:
\begin{flalign}\label{y_est}
\boldsymbol{Y_p}=\sqrt{\rho_p} {\boldsymbol{C}\boldsymbol{\Phi}}+\boldsymbol{N} \ \ 
\end{flalign}
where $\boldsymbol{Y_p}$ and $\boldsymbol{N}$ are $M \times N_p$ matrices of received samples and noise samples, respectively, and $\boldsymbol{C}$ is equal to $\boldsymbol{H}\boldsymbol{D}^{1/2}$ where $\boldsymbol{H}$ is the $M \times K$ matrix of fading coefficients between the K users and the BS, i.e., $\boldsymbol{H}(k,m)=h_{km}$, and $\boldsymbol{D}$ is a $K \times K$ diagonal matrix containing the path-loss coefficients, i.e., $\boldsymbol{D}(k,k)=\beta_k$. Also, $\rho_p$ is the power assigned to transmission of pilot sequences and is equal to $\rho_p=N_p \rho_d$. The least squares estimate of the channel matrix $\boldsymbol{C}$ can be calculated as:
\begin{flalign}
\tilde{\boldsymbol{C}}&=\frac{1}{\sqrt {\rho_p}}\boldsymbol{Y_p} \boldsymbol{\Phi}^H 
\end{flalign}
Then, $\tilde{\boldsymbol{C}}$ is used to perform MRC. However, due to the existence of unknown time delays, the estimation process is degraded and as a result, the channel estimations are contaminated by unwanted channel coefficients. In what follows, we provide a similar analysis for the channel estimation when the misalignment exists between received signals. Time delays modify Eq. (\ref{y_est}) to:
\begin{flalign}
\boldsymbol{Y_p}=\sqrt{\rho_p} \sum_{i=-I}^{I}{\boldsymbol{C^i}\boldsymbol{\Phi^i}}+\boldsymbol{N} 
\end{flalign}
where $\boldsymbol{C^i}_{M\times K}$ and $\boldsymbol{\Phi^i}_{K\times N_p}$ are defined as follows:
\small
\begin{flalign}
\nonumber
\boldsymbol{C^i}&=\begin{pmatrix}
g(eT_s+T+iT_s-\tau_{11})\sqrt{\beta_1}h_{11}&\cdots & g(eT_s+T+iT_s-\tau_{K1})\sqrt{\beta_K}h_{K1}\\ 
g(eT_s+T+iT_s-\tau_{12})\sqrt{\beta_1}h_{12}&\cdots & g(eT_s+T+iT_s-\tau_{K2})\sqrt{\beta_K}h_{K2} \\ 
\vdots & \ddots & \vdots\\ 
g(eT_s+T+iT_s-\tau_{1M})\sqrt{\beta_1}h_{1M}&\cdots & g(eT_s+T+iT_s-\tau_{KM})\sqrt{\beta_K}h_{KM}
\end{pmatrix}
\end{flalign}
\begin{flalign}
\nonumber
\boldsymbol{\Phi^{i\leq 0}}=\begin{pmatrix}
p_1(1-i)&\cdots & p_1(N_p)&0&\cdots&0\\ 
p_2(1-i)&\cdots & p_2(N_p) &0&\cdots&0\\ 
\vdots & \ddots & \vdots\\ 
p_K(1-i)&\cdots & p_K(N_p)&0&\cdots&0
\end{pmatrix}, 
\boldsymbol{\Phi^{i\geq 0}}=\begin{pmatrix}
0&\cdots&0& p_1(1)&\cdots & p_1(N_p-i)\\ 
0&\cdots&0&p_2(1)&\cdots & p_2(N_p-i) \\ 
\vdots & \ddots & \vdots\\ 
0&\cdots&0&p_K(1)&\cdots & p_K(N_p-i)
\end{pmatrix}
\end{flalign} 
\normalsize
The process of de-spreading, which is multiplying the received pilot signal by $\frac{1}{\sqrt{\rho_p}}\boldsymbol{\Phi}^H$, yields the following channel matrix estimator:
\begin{flalign}
\tilde{\boldsymbol{C}}=\sum_{\substack{i=-I}}^{I}{\boldsymbol{C^i}\boldsymbol{\Phi^i}\boldsymbol{\Phi}^H}+\tilde{\boldsymbol{N}}
\end{flalign}
where $\tilde{\boldsymbol{N}}$ is the estimation noise. We denote $\boldsymbol{\Phi^i}\boldsymbol{\Phi}^H$ by $\boldsymbol{\Upsilon^i}$ which is equal to $\boldsymbol{I_K}$ for $i=0$, and for the other values of $i$ can be calculated as:
\small
\begin{flalign}
\nonumber
(\boldsymbol{\Upsilon^{i<0}})^T=\boldsymbol{\Upsilon^{i>0}}=\begin{pmatrix}
\langle\ \boldsymbol{p_1}(1:N_p-i).{\boldsymbol{p_1}(1+i:N_p)}\rangle&\cdots & \langle\ \boldsymbol{p_1}(1:N_p-i).{\boldsymbol{p_K}(1+i:N_p)}\rangle\\ 
\vdots & \ddots & \vdots\\ 
\langle\ \boldsymbol{p_K}(1:N_p-i).{\boldsymbol{p_1}(1+i:N_p)}\rangle&\cdots & \langle\ \boldsymbol{p_K}(1:N_p-i).{\boldsymbol{p_K}(1+i:N_p)}\rangle
\end{pmatrix}
\end{flalign}
\normalsize
where $\boldsymbol{p}(i:j)$ represents the vector $[p(i), p(i+1), \cdots, p(j)]$. After some calculations, estimate of the channel coefficient of User $l$ to receive antenna $m$ can be represented as:
\begin{flalign}
\tilde{c}_{lm}=\sum_{\substack{j=1}}^{K}{\lambda_{ljm}}c_{jm}+\tilde{n}_{lm}
\label{jadid}
\end{flalign}
where $\lambda_{ljm}$ is the leakage from User $j$ to the estimation of the User $l$'s channel coefficient to receive antenna $m$ and is equal to:
\begin{flalign}
\lambda_{ljm}=\sum_{\substack{i=-I}}^{I}{g(eT_s+T+iT_s-\tau_{jm})\boldsymbol{\Upsilon^i}(j,l)}
\end{flalign}
This phenomenon is similar to the ``pilot contamination" effect, i.e., the channel estimation of each user to the $m$th receive antenna is contaminated by channel coefficients of other users. In the ``pilot contamination" problem, the reason of contamination is reusing the same pilot sequences in different cells, however, here, the reason is unknown timing mismatches between the received signals. Due to the time asynchrony between the received signals, the orthogonality between pilot sequences is not preserved anymore and the de-spreading matrix is not able to eliminate the effect of interfering users. Hence, designing pilot sequences with good properties like low cross correlations can decrease the contamination. Particularly, the Zadoff-Chu (ZC) sequence exhibits the useful property that cyclically shifted versions of the sequences are orthogonal to one another. By using these sequences in the estimation phase, we can eliminate the contamination caused by the loss of orthogonality and channel estimates turn into $\tilde{c}_{lm}=c_{lm}+\tilde{n}_{lm}$. Therefore, thanks to the ZC sequences, IUI is avoided by preserving the orthogonality in the estimation process; however, ISI, caused from imperfect sampling, is still degrading the performance as will be shown in Eq. (\ref{far4}). \\

In general, for any contaminated channel estimates presented as Eq. (\ref{jadid}), the effective channel matrix and noise vector after MRC can be expressed, respectively, as:
\small
\begin {flalign}\label{effective_matrix_ip}
\boldsymbol{T^{mrc}_{lk,ip}}=\frac{1}{M}\sum_{m=1}^{M}\sum_{j=1}^{K}\lambda_{ljm}\sqrt{\beta_j \beta_k}h_{jm}^{\ast}h_{km}\boldsymbol{G_{km}}
\end{flalign}
\begin{flalign}\label{noise_ip}
\boldsymbol{n_{l,ip}^{mrc}}=\frac{\sqrt{\rho_d}}{M}\sum_{m=1}^{M}\tilde n_{lm} \sum_{k=1}^{K}\sqrt{\beta_k}h_{km}\boldsymbol{G_{km}}\boldsymbol{b_k}+\frac{1}{M}\sum_{m=1}^{M}\left(\sum_{j=1}^{K}\lambda_{ljm}\sqrt{\beta_j }h_{jm}^{\ast}+\tilde n_{lm}\right)\boldsymbol{n_m}
\end{flalign}
\normalsize
The corresponding achievable rate results are presented in the next theorem. 
\begin{theorem}{\label{22}}
The achievable rate by the MRC receiver using orthogonal channel estimation, when there is unknown time delays between received signals can be approximated as follows:
\begin{flalign}\label{ss}
{R}_{l,ip}^{mrc}\approx\kappa\log_{2}{\left(1+\frac{\rho_d \beta_l^2M(\gamma'_{lll}(0))^2}{IUI+ISI+noise}\right)}
\end{flalign}
where $\kappa=\frac{N-N_p}{N}$ accounts for the spectral efficiency loss due to channel estimation, and ISI, IUI and noise components are defined, respectively, as follows:
\small
\begin{flalign}
\nonumber
ISI&=\rho_d \beta_l^2\sum\limits_{\substack{n=-I}}^{I}{(2\gamma''_{lll}(n)+(M\bar\delta[n]-1)(\gamma'_{lll}(n))^2)}+\rho_d\beta_l\sum\limits_{\substack{j=1\\j\neq l}}^{K}{\beta_j\sum\limits_{\substack{n=-I}}^{I}\gamma''_{ljl}(n)}\\
\nonumber
IUI&=\rho_d \sum\limits_{\substack{k=1\\k\neq l}}^{K}\beta_k^2\sum\limits_{\substack{n=-I}}^{I}{(2\gamma''_{lkk}(n)+(M-1)(\gamma'_{lkk}(n))^2)}+\rho_d\sum\limits_{\substack{k=1\\k\neq l}}^K\sum\limits_{\substack{j=1\\j\neq k}}^{K}{\beta_{k}\beta_{j}\sum\limits_{\substack{n=-I}}^{I}\gamma''_{ljk}(n)}\\
\nonumber
noise&=\frac{\rho_d}{\rho_p}\sum\limits_{\substack{k=1}}^{K}{\beta_k\sum\limits_{\substack{n=-I}}^{I}E[g^2_n]}+\sum_{k=1}^{K}{\beta_{k}\lambda''_{lk}}+\frac{1}{\rho_p}
\end{flalign}
\normalsize
where
\small
\begin{flalign}
\nonumber
\gamma'_{ljk}(n)&=E[\gamma_{ljkm}(n)]=E[\lambda_{ljm}g(eT_s+T+nT_s-\tau_{km})]\\
\label{baad}
&=\int_{-\infty}^{\infty}\int_{-\infty}^{\infty}{\left( \sum_{\substack{i=-I}}^{I}{\boldsymbol{\Upsilon^i}(j,l)g(eT_s+T+iT_s-\tau_{j})}\right)g(eT_s+T+nT_s-\tau_k)f(\tau_j)f(\tau_k)d\tau_jd\tau_k}
\end{flalign}
\normalsize
Assuming the same distribution for all time delays, the receive antenna index is discarded after taking expectations. The terms $\gamma''_{ljk}(n)$ and $\lambda_{lk}^{''}$ are defined similarly as:
\small
\begin{flalign}
\nonumber
\gamma''_{ljk}(n)=\int_{-\infty}^{\infty}\int_{-\infty}^{\infty}{\left( \sum_{\substack{i=-I}}^{I}{\boldsymbol{\Upsilon^i}(j,l)g(eT_s+T+iT_s-\tau_{j})}\right)^2g^2(eT_s+T+nT_s-\tau_k)f(\tau_j)f(\tau_k)d\tau_jd\tau_k}
\end{flalign}
\begin{flalign}
\nonumber
\lambda''_{lk}=E[\lambda^2_{lkm}]=\int_{-\infty}^{\infty}{\left( \sum_{\substack{i=-I}}^{I}{\boldsymbol{\Upsilon^i}(k,l)g(eT_s+T+iT_s-\tau_{k})}\right)^2f(\tau_k)d\tau_k}
\end{flalign}
\normalsize
\end{theorem}
\begin{IEEEproof}
The proof is presented in Appendix \ref{appendix_2}. 
\end{IEEEproof}
These results are general for any pilot matrices and delay distributions. Values of $\gamma'_{ljk}(n), \gamma''_{ljk}(n), \lambda''_{lk}$ and $E[g^2_n]$ only depend on the pulse shape, pilot sequences and the delay distribution which can be calculated analytically or numerically. Thus, we can insert any delay distribution, any pilot matrix and any value of $K$ in the theorem and calculate the achievable rates. For example, in the case of using ZC sequences where the contamination in the estimation process is eliminated, the values of $\gamma'_{ljk}(n), \gamma''_{ljk}(n), \lambda''_{lk}$ are equal to $E[g_n]\delta[l-j]$, $E[g^2_n]\delta[l-j]$ and $\delta[l-k]$, respectively. Then, the achievable rate will be equal to: 

\begin{flalign}
\label{far4}
R_{l,p}^{mrc-ZC}\approx
\kappa\log_{2}\biggl(1+
  \frac{N_p\rho^2_d \beta_l^2ME^2[g_0]}{\Delta+1}\biggr)
\end{flalign}
where $\Delta$ is defined as:

\small
\begin{flalign}
\nonumber
\Delta=\rho_d(N_p\rho_d\beta_l+1)\sum\limits_{i=-I}^{I}{E[g_i^2]}\sum\limits_{\substack{k=1 \\ k\neq l}}^{K}{\beta_k}+N_p\rho^2_d \beta^2_l\sum\limits_{\substack{i=-I }}^{I}(2E[g_i^2]+(M\bar\delta[i]-1)E^2[g_i])
+\rho_d\left(N_p+\sum\limits_{i=-I}^{I}{E[g_i^2]}\right)\beta_l
\end{flalign}
\normalsize

The ideal synchronized case admits a power-scaling in the order of $\frac{1}{\sqrt{M}}$ with no degradation in the achievable rate of each user, i,e., ${R}_{l,ip}^{ideal}\rightarrow\log_{2}{(1+N_p E_d^2\beta_l^2)}\ $ as $\ M\rightarrow \infty, \rho_d=\frac{E_d}{\sqrt{M}}$. \cite{ngo2013energy}. However, due to the existence of timing mismatch, the promised power scaling law is lost. If we reduce the transmit power by an order of $\frac{1}{\sqrt{M}}$ and let $M$ go to infinity, then the achievable rate for each user is:
\small
\begin{flalign}
R_{l,ip}^{mrc}\rightarrow\kappa\log_{2}{\left(1+\frac{N_pE_d^2\beta_l^2(\gamma'_{lll}(0))^2}{N_pE_d^2\beta_l^2\sum\limits_{\substack{i=-I\\i \neq 0}}^{I}(\gamma'_{lll}(i))^2+N_pE_d^2\sum\limits_{\substack{k=1\\k\neq l}}^{K}\beta_k^2\sum\limits_{\substack{i=-I}}^{I}(\gamma'_{lkk}(i))^2+1}\right)}\ 
\end{flalign}
\normalsize
By increasing the transmit power, the achievable rate saturates at the following fixed value:
\small
\begin{flalign}{\label{salam}}
R_{l,ip}^{mrc}\rightarrow\kappa\log_{2}{\left(1+\frac{\beta^2_l(\gamma'_{lll}(0))^2}{\beta_l^2\sum\limits_{\substack{i=-I\\i \neq 0}}^{I}(\gamma'_{lll}(i))^2+\sum\limits_{\substack{k=1\\k\neq l}}^{K}\beta_k^2\sum\limits_{\substack{i=-I}}^{I}(\gamma'_{lkk}(i))^2}\right)}\ 
\end{flalign}
\normalsize
The above analysis shows that, the promised single-user bound is degraded by two factors: ISI (due to adjacent symbols of the desired user) and IUI (due to contamination in the estimation process). For any given pulse shape and pilot sequence, the performance in Eq. (\ref{salam}) can be optimized by changing the value of $e$. For example, the optimum value of $e$ for the rectangular pulse shape and the delay distribution described in Eq. (\ref{delays}), is presented in Table \ref{table2}.
\begin{table}[h!]
\caption{Optimal Sampling Origin $e$} 
\centering 
\begin{tabu}{c c c c c c c c c}
\hline\hline 
Case & $K=2$ & $K=4$ & $K=6$ & $K=8$ & $K=10$ & $K=12$ & $K=14$ & $K=16$ \\ 
\hline 
Optimal $e$ & 0 & 0.12 & 0.28& 0.38 & 0.44 & 0.46 & 0.46 & 0.48\\ 
\hline 
\end{tabu}
\label{table2}
\end{table}
Similar to the perfect CSI case, as $K$ increases, the optimal value of $e$ approaches half. 
While such an optimization can increase the achievable rates in Theorems 1 and 2, the main benefit of massive MIMO setting which is unlimited achievable rate by using asymptotically large number of receive antennas will still be out of reach in the presence of unknown time delays. Therefore, we design two receiver structures, for perfect CSI and imperfect CSI scenarios, to remove the unwanted effects of ISI and IUI imposed by unknown time delays. The details are presented next.

\section{the Achievable Rate of MRC-ZF receiver}\label{sec-MRCZF}

In an ideal massive MIMO system, the achievable uplink rate grows unbounded when $M$ grows large. Therefore, we can scale down the power of each user by the ratios of $M$ and $\sqrt{M}$ for the perfect and imperfect CSI, respectively, to achieve the single-user performance with no interference. However, in a realistic massive MIMO system where timing mismatch among received signals is inevitable, the uplink achievable rate using the MRC receiver approaches a constant value when $M$ grows large, as shown in Theorems \ref{11} and \ref{22}. Therefore, the power scaling law, provided by the MRC receiver in a perfectly synchronized massive MIMO system, is not achievable with unknown time delays. To understand the underlying reason, we express the expected value of the effective channel matrices obtained by the MRC receiver, i.e., $\boldsymbol{T^{mrc}_{lk,p}}$ and $\boldsymbol{T^{mrc}_{lk,ip}}$, respectively:
\small
\begin {flalign}{\label{qwer}}
E[{\boldsymbol{T^{mrc}_{lk,p}}}]=\sqrt{\beta_l \beta_k}\delta{[l-k]}\begin{pmatrix}
{E}[g_{0}]&{E}[g_{-1}]&\cdots & {E}[g_{1-N}]\\ 
{E}[g_{1}]&{E}[g_{0}]& \cdots & {E}[g_{2-N}]\\ 
\vdots & \ddots &\ddots& \vdots\\ 
{E}[g_{N-1}]&{E}[g_{N-2}]&\cdots &{E}[g_{0}]
\end{pmatrix}
\end{flalign}
\begin {flalign}
 E[\boldsymbol{T^{mrc}_{lk,ip}}]=\sum_{j=1}^{K}\delta [j-k]\sqrt {\beta _j\beta_k}E[\lambda_{ljm}\boldsymbol{G_{km}}]=\beta_k\begin{pmatrix}{\label{qwer2}}
\gamma'_{lkk}(0)&\gamma'_{lkk}(-1)&\cdots & \gamma'_{lkk}(1-N)\\ 
\gamma'_{lkk}(1)&\gamma'_{lkk}(0)& \cdots & \gamma'_{lkk}(2-N)\\ 
\vdots & \ddots &\ddots& \vdots\\ 
\gamma'_{lkk}(N-1)&\gamma'_{lkk}(N-2)&\cdots &\gamma'_{lkk}(0)
\end{pmatrix}
\end{flalign}
\normalsize

Based on the law of large numbers, when M grows large, the effective channel matrices approach their expected values. Therefore, in the perfect CSI case, where the expected value of the effective channel matrices from unwanted users is zero, the dominant degradation will be ISI as $M$ increases. On the other hand, in the imperfect CSI case, the expected value of the effective channel matrices is nonzero for all the users, thus, not only ISI but also IUI will degrade the performance. 

\begin{figure*}[!t]
\centering
\subfloat[MRC-ZF Receiver with perfect CSI]{\includegraphics[width=6in]{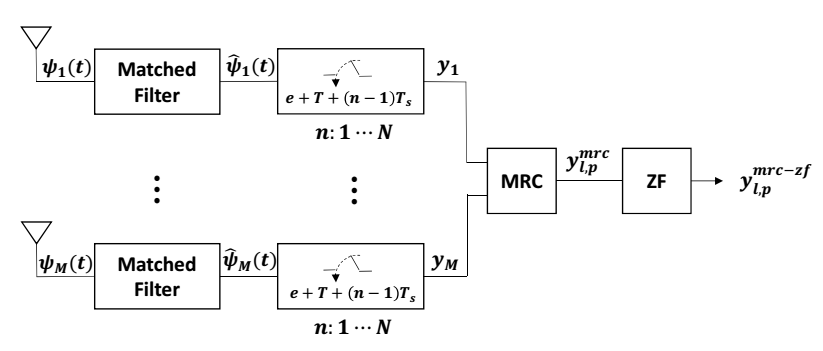}
\label{fig:system_mrc_zf_p}}
\hfil
\subfloat[MRC-ZF Receiver with imperfect CSI]{\includegraphics[width=6in]{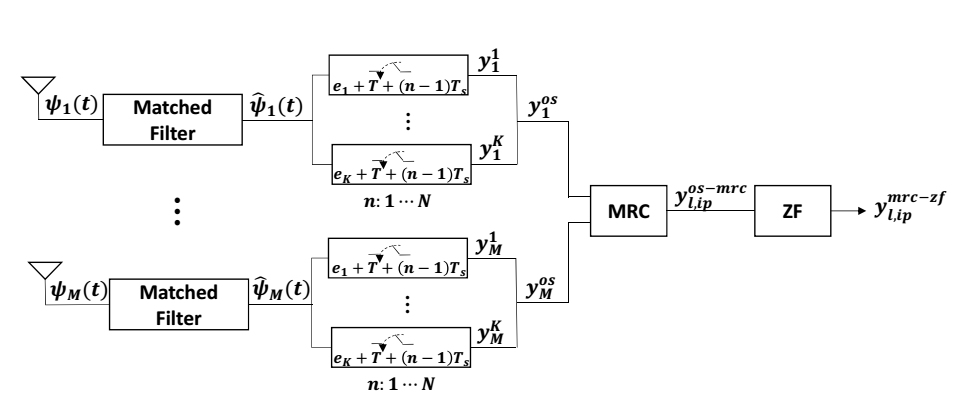}
\label{fig:system_mrc_zf_ip}}
\caption{MRC-ZF Receiver}
\end{figure*}

To cancel the effect of these impairments, we use the concept of zero forcing (ZF), however, other methods like minimum mean squared error (MMSE) and successive interference cancellation (SIC) can also be used. For the perfect CSI scenario, where only the desired user's effective channel matrix has a nonzero expectation, its effect can be canceled by multiplying the output sample of the MRC receiver by matrix $\boldsymbol{Z}$ defined as:
\small
\begin {flalign}\label{Z}
\boldsymbol Z=\begin{pmatrix}
{E}[g_{0}]&{E}[g_{-1}]&\cdots & {E}[g_{1-N}]\\ 
{E}[g_{1}]&{E}[g_{0}]& \cdots & {E}[g_{2-N}]\\ 
\vdots & \ddots &\ddots& \vdots\\ 
{E}[g_{N-1}]&{E}[g_{N-2}]&\cdots &{E}[g_{0}]
\end{pmatrix}^{-1}
\end{flalign}
\normalsize
For the imperfect CSI scenario, because all users' effective channel matrices have nonzero expectations, we need extra sets of equations to cancel all the interference terms. Therefore, we utilize the concept of oversampling as explained in \cite{poorkasmaei2015asynchronous,avendi2015differential,mehdi}. The receiver structure for the proposed methods are shown in Figures \subref*{fig:system_mrc_zf_p} and \subref*{fig:system_mrc_zf_ip}. The description of these methods follows next.  

\subsection{Perfect CSI}
As shown in Fig. \ref{fig:system_mrc_zf_p}, the output samples of the MRC receiver are multiplied by $\boldsymbol{Z}$ to cancel the effect of the averaged ISI. Note that matrix $\boldsymbol{Z}$ is pre-calculated once based on the pulse shape, sampling origin and delay distributions and then, it can be used during the entire transmission. We call this receiver MRC-ZF whose output samples are:
\small
\begin{flalign}\label{sys_mrczf}
\boldsymbol{{y}_{l,p}^{mrc-zf}}&=\sqrt{\rho_d} \sum_{k=1}^{K}{\boldsymbol{T^{mrc-zf}_{lk,p}}\boldsymbol{b_k}}+\boldsymbol{ n^{mrc-zf}_{l,p}}
\end{flalign}
\normalsize
where $\boldsymbol{T^{mrc-zf}_{lk,p}}=\boldsymbol{Z T^{mrc}_{lk,p}}$ and $\boldsymbol{ n^{mrc-zf}_{l,p}}=\boldsymbol{Z n^{mrc}_{l,p}}$. For the special case of symbol-level synchronization, i.e., $f(\tau)=\delta(\tau)$, $\boldsymbol{Z}$ will be the identity matrix; meaning that no additional processing is required as done in the literature. The expected value of matrix $\boldsymbol{T^{mrc-zf}_{lk,p}}$ is equal to $\sqrt{\beta_l\beta_k}\delta[l-k]\boldsymbol{I_N}$, which means the effect of ISI diminishes for large values of M. The approximation of the achievable rate by the MRC-ZF receiver is presented in the next theorem.
\begin{theorem}\label{33}
The achievable rate by each user using the MRC-ZF receiver can be approximated by:
\small
\begin{flalign}\label{coffe_bean}
R_{l,p}^{mrc-zf}\approx\log_{2}{\left(1+\frac{\rho_d \beta_lM}{\rho_d(\sum_{\substack{n=0}}^{N-1}\xi''_n)\sum\limits_{\substack{k=1\\k\neq l}}^{K}{\beta_k}+\rho_d \beta_l(2\sum_{\substack{n=0\\}}^{N-1}\xi''_{n}-1)+\epsilon_0}\right)}
\end{flalign}
\normalsize
where $\xi''_{p-q}=E\left[\boldsymbol{G'}^2_{\boldsymbol{}}(p,q)\right]$, $\boldsymbol{G'}=\boldsymbol{ZG}$ and $\epsilon_{p-q}=\boldsymbol{Z}\boldsymbol{Z}^H(p,q)$ are only functions of the distribution of delays and the pulse shape. Assuming the same distribution for all the time delays, receive antenna and
user indexes are discarded after taking expectations. Note that, throughout the paper, for element-wise power we use $\left(\boldsymbol{G}_{\boldsymbol{}}(p,q)\right)^2$ and for elements of powered matrices we use $\boldsymbol{G}^2_{\boldsymbol{}}(p,q)$.

\end{theorem}
\begin{IEEEproof}
The proof is presented in Appendix \ref{appendix_3}. 
\end{IEEEproof}
By using the MRC-ZF receiver which exploits the statistics of the unknown time delays, the effect of the averaged ISI is vanished. If the number of receive antennas goes to infinity, the achievable rate goes to infinity. Therefore, we can scale down the transmit power and still provide the desirable performance. In other words, if we choose $\rho_d=\frac{E_d}{M}$ in Eq. (\ref{coffe_bean}) and let $M$ go to infinity, we will have:
\small
\begin{flalign}
R_{l,p}^{mrc-zf}\rightarrow\log_{2}{\left(1+\frac{E_d\beta_l}{\epsilon_0}\right)},  \textmd{as} \ M\rightarrow \infty,  \rho_d=\frac{E_d}{M}
\end{flalign}
\normalsize
Hence, even in the presence of unknown time delays, the power scaling law is held for the MRC-ZF receiver. The value of $\epsilon_0$ is calculated based on the pulse shape, time delay distribution and sampling origin. The loss of $\epsilon_0$ is because of the noise enhancement by ZF and can be mitigated by using other cancellation methods like MMSE and SIC.
\subsection{Imperfect CSI} \label{pool}
As explained before, when the channel coefficients are estimated by using orthogonal pilot sequences, the effective channel matrices for all the $K$ users are nonzero. Therefore, we need at least $K$ sets of samples to cancel them. Denoting $\boldsymbol{y_m^t}, 1\leq t \leq K$ as the set of $N$ samples, obtained at sampling times of $e_tT_s+T+(n-1)T_s, n=1,\cdots, N$, we collect all the samples obtained at receive antenna $m$ in a vector $\boldsymbol{y^{os}_m}=[(\boldsymbol{y^1_m})^T,\cdots, (\boldsymbol{y^K_m})^T]^T$ to derive:
\begin{flalign}
\boldsymbol{y^{os}_m}=\sqrt{\rho_d}\boldsymbol{T_m}\boldsymbol{b}+\boldsymbol{n^{os}_m}
\end{flalign}
where $\boldsymbol{b}=[\boldsymbol{b_1}^T,\cdots,\boldsymbol{b_K}^T]^T$ includes transmitted vectors of all users and $\boldsymbol{T_m}$ is defined as:
\small
\begin{flalign}
\boldsymbol{T_m}=\begin{pmatrix}
\boldsymbol{T_{1m}^1}& \boldsymbol{T_{2m}^1}&\cdots & \boldsymbol{T_{Km}^1}\\ 
\boldsymbol{T_{1m}^2}& \boldsymbol{T_{2m}^2}& \cdots & \boldsymbol{T_{Km}^2}\\ 
\vdots & \ddots &\ddots& \vdots\\ 
\boldsymbol{T_{1m}^K}& \boldsymbol{T_{2m}^K}&\cdots & \boldsymbol{T_{Km}^K}
\end{pmatrix}
\end{flalign}
\normalsize
where $\boldsymbol{T_{km}^t}$ represents the channel matrix of User $k$ to receive antenna $m$ in the $t$th set of samples, i.e., $\boldsymbol{T^t_{km}}=\sqrt{\beta_k}h_{km}\boldsymbol{G_{km}^t}$ where $\boldsymbol{G_{km}^t}$ is defined by Eq. (\ref{gg}) with $e=e_t$.
The noise vector also includes all the noise vectors obtained from different sampling times, i.e., $\boldsymbol{n^{os}_m}=[(\boldsymbol{n^1_m})^T,\cdots,(\boldsymbol{n^K_m})^T]^T$ and its covariance matrix is calculated by:
\small
\begin{flalign}
\boldsymbol{\Sigma_{n^{os}}}=\begin{pmatrix}
\boldsymbol{\Sigma_{11}}&\boldsymbol{\Sigma_{12}}&\dots & \boldsymbol{\Sigma_{1K}}\\ 
\boldsymbol{\Sigma_{21}} & \boldsymbol{\Sigma_{22}}&\dots &\boldsymbol{\Sigma_{2K}}\\ 
\vdots&\ddots & \ddots &\vdots\\ 
\boldsymbol{\Sigma_{K1}}&\boldsymbol{\Sigma_{K2}}&\dots&\boldsymbol{\Sigma_{KK}}
\end{pmatrix}
\end{flalign}
\normalsize
where $\boldsymbol{\Sigma_{t_1t_2}}$ is the covariance matrix between the noise samples obtained at times $t_1$ and $t_2$:
\small
\begin {flalign}
\boldsymbol{\Sigma_{t_1,t_2}}=\begin{pmatrix}
g(T+(e_{t_1}-e_{t_2})T_s)&\cdots & g(T+(1-N)T_s+(e_{t_1}-e_{t_2})T_s)\\ 
g(T+T_s+(e_{t_1}-e_{t_2})T_s)& \cdots & g(T+(2-N)T_s-(e_{t_1}-e_{t_2})T_s) \\ 
\vdots & \ddots & \vdots\\ 
g(T+(N-1)T_s+(e_{t_1}-e_{t_2})T_s)&\cdots & g(T+(e_{t_1}-e_{t_2})T_s)
\end{pmatrix}
\end{flalign}
\normalsize
The receive antenna index is discarded because the noise covariance matrix is the same at all receive antennas. The channel coefficient of User $l$ to receive antenna $m$ , i.e., $\tilde{c}^s_{lm}$ is equal to:
\begin{flalign}
\tilde{c}^s_{lm}=\sum_{\substack{j=1}}^{K}{\lambda^s_{ljm}}c_{jm}+\tilde{n}^s_{lm}
\end{flalign}
where $\lambda^s_{ljm}$ is equal to:
\begin{flalign}
\lambda^s_{ljm}=\sum_{\substack{i=-I}}^{I}{g(e_sT_s+T+iT_s-\tau_{jm})\boldsymbol{\Upsilon^i}(j,l)}
\end{flalign}
Because oversampling is only performed in the data detection phase, we use index $s$ for the set of sampling times in the channel estimation phase to differentiate it from the sets of
sampling times in the data detection phase. After performing MRC for the $l$th user, the resulting system of equations is:
\begin{flalign}
\boldsymbol{{y}_{l,ip}^{os-mrc}}&=\sqrt{\rho_d} \boldsymbol{\hat T_{l}}\boldsymbol{b}+\boldsymbol{ n^{os-mrc}_{l,ip}}
\end{flalign}
where $\boldsymbol{\hat T_{l}}$ is the effective channel matrix:
\small
\begin{flalign}
\boldsymbol{\hat T_{l}}=\begin{pmatrix}
\boldsymbol{\hat T_{l1}^1}& \boldsymbol{\hat T_{l2}^1}&\cdots & \boldsymbol{\hat T_{lK}^1}\\ 
\boldsymbol{\hat T_{l1}^2}& \boldsymbol{\hat T_{l2}^2}& \cdots & \boldsymbol{\hat T_{lK}^2}\\ 
\vdots & \ddots &\ddots& \vdots\\ 
\boldsymbol{\hat T_{l1}^K}& \boldsymbol{\hat T_{l2}^K}&\cdots & \boldsymbol{\hat T_{lK}^K}
\end{pmatrix}
\end{flalign}
\normalsize
and each sub-block, $\boldsymbol{\hat T_{lk}^t}$, is defined as:
\small
\begin{flalign}
\boldsymbol{\hat T^t_{lk}}=\frac{1}{M}\sum_{m=1}^{M}\left(\sum_{\substack{j=1}}^{K}{\lambda^s_{ljm}}c^*_{jm}\right)\boldsymbol{T_{km}^t}
\end{flalign}
\normalsize
The effective noise is also calculated as:
\small
\begin{flalign}
\boldsymbol{n_{l,ip}^{os-mrc}}=\frac{\sqrt{\rho_d}}{M}\sum_{m=1}^{M}(\tilde n^{s}_{lm})^* \boldsymbol{T_{m}}\boldsymbol{b}+\frac{1}{M}\sum_{m=1}^{M}\left(\sum_{\substack{j=1}}^{K}{\lambda^s_{ljm}}c^*_{jm}+(\tilde{n}^s_{lm})^*\right)\boldsymbol{n^{os}_m}
\end{flalign}
\normalsize
Using Eq. (\ref{qwer2}), the expected value of $\boldsymbol{\hat T_{l}}$ is equal to:
\small
\begin{flalign}
\label{71}
E[\boldsymbol{\hat T_{l}}]&=\begin{pmatrix}
\boldsymbol{\Gamma_{l1}^1}& \boldsymbol{\Gamma_{l2}^1}&\cdots & \boldsymbol{\Gamma_{lK}^1}\\ 
\boldsymbol{\Gamma_{l1}^2}& \boldsymbol{\Gamma_{l2}^2}& \cdots & \boldsymbol{\Gamma_{lK}^2}\\ 
\vdots & \ddots &\ddots& \vdots\\ 
\boldsymbol{\Gamma_{l1}^K}& \boldsymbol{\Gamma_{l2}^K}&\cdots & \boldsymbol{\Gamma_{lK}^K}
\end{pmatrix}\begin{pmatrix}
\beta_1\boldsymbol{I_N}& \boldsymbol{0}&\cdots & \boldsymbol{0}\\ 
\boldsymbol{0}& \beta_2\boldsymbol{I_N}& \cdots & \boldsymbol{0}\\ 
\vdots & \ddots &\ddots& \vdots\\ 
\boldsymbol{0}& \boldsymbol{0}&\cdots & \beta_K\boldsymbol{I_N}
\end{pmatrix}=\boldsymbol{\Gamma_l}\begin{pmatrix}
\beta_1\boldsymbol{I_N}& \boldsymbol{0}&\cdots & \boldsymbol{0}\\ 
\boldsymbol{0}& \beta_2\boldsymbol{I_N}& \cdots & \boldsymbol{0}\\ 
\vdots & \ddots &\ddots& \vdots\\ 
\boldsymbol{0}& \boldsymbol{0}&\cdots & \beta_K\boldsymbol{I_N}
\end{pmatrix}
\end{flalign}
\normalsize
where $\boldsymbol{\Gamma_{lk}^t}$ is calculated as:
\small
\begin {flalign}
\boldsymbol{\Gamma_{lk}^t}=\begin{pmatrix}
\gamma'_{lkkt}(0)&\gamma'_{lkkt}(-1)&\cdots & \gamma'_{lkkt}(1-N)\\ 
\gamma'_{lkkt}(1)&\gamma'_{lkkt}(0)& \cdots & \gamma'_{lkkt}(2-N)\\ 
\vdots & \ddots &\ddots& \vdots\\ 
\gamma'_{lkkt}(N-1)&\gamma'_{lkk}(N-2)&\cdots &\gamma'_{lkkt}(0)
\end{pmatrix}
\end{flalign}
\normalsize
and its elements are equal to:
\small
\begin{flalign}
\nonumber
\gamma'_{lkkt}(n)&=E[\lambda^s_{lkm}g(e_tT_s+T+nT_s-\tau_{km})]\\
&=\int_{-\infty}^{\infty}{\left( \sum_{\substack{i=-I}}^{I}{\boldsymbol{\Upsilon^i}(k,l)g(e_sT_s+T+iT_s-\tau_{k})}\right)g(e_tT_s+T+nT_s-\tau_k)f(\tau_k)d\tau_k}
\end{flalign}
\normalsize
Note that values of $\gamma'_{ljkt}(n)$ are basically defined the same as $\gamma'_{ljk}(n)$ in Eq. \ref{baad}; however, because
of oversampling, there is an extra index of $t$ which represents the sampling origin index, i.e., $t$
in $e_t$. 

Matrix $\boldsymbol{\Gamma_l}$ in Eq. (\ref{71}) is only related to sampling origins, i.e., $e_t$s, pilot sequences, the pulse shape and delay distributions and is known at the receiver. To resolve the problem of ISI and IUI, we calculate the inverse of $\boldsymbol{\Gamma_l}$ and denote it as $\boldsymbol{W_l}$, which is constructed by sub-blocks of $\boldsymbol{W_{lk}}$, i.e., $\boldsymbol{W_l}=[\boldsymbol{W_{l1}}^T,\cdots, \boldsymbol{W_{lK}}^T]^T$. Then, in order to detect the transmitted symbols of the $l$th user, we multiply the output of the MRC receiver by the $l$th sub-block of $\boldsymbol{W_l}$, i.e., $\boldsymbol{W_{ll}}$. Therefore, the resulting samples will be:
\begin{flalign}\label{milk_choc}
\boldsymbol{y_{l,ip}^{mrc-zf}}=\sqrt{\rho_d}\sum_{k=1}^{K}{\boldsymbol{T_{lk,ip}^{mrc-zf}}\boldsymbol{b_k}+\boldsymbol{n_{l,ip}^{mrc-zf}}}
\end{flalign}
where $\boldsymbol{T_{lk,ip}^{mrc-zf}}=\boldsymbol{W_{ll}\hat T_{lk}}$, $\boldsymbol{n_{l,ip}^{mrc-zf}}=\boldsymbol{W_{ll} n^{os-mrc}_{l,ip}}$ and $\boldsymbol{\hat T_{lk}}=[(\boldsymbol{T^1_{lk}})^T,\dots,(\boldsymbol{T^K_{lk}})^T]^T$. It can be shown that, the expected value of matrix $\boldsymbol{T^{mrc-zf}_{lk,ip}}$ is equal to $\sqrt{\beta_l\beta_k}\delta[l-k]\boldsymbol{I_N}$. Therefore, as $M$ grows large, $\boldsymbol{T^{mrc-zf}_{lk,ip}}$ will be closer to its expected value and the effect of ISI and IUI converges to zero. The achievable rates for the aforementioned system is presented in the next theorem.
\begin{theorem}\label{44}
The achievable rate by the MRC-ZF receiver using estimated channel coefficients,
when there is unknown time delays between the received signals can be approximated by:
\begin{flalign}\label{mrc_zf2}
{R}_{l,ip}^{mrc-zf}\approx\kappa\log_{2}{\left(1+\frac{\rho_d \beta_l^2M}{IUI+ISI+noise}\right)}
\end{flalign}
where ISI, IUI and noise components are defined, respectively, as:
\small
\begin{flalign}
\nonumber
ISI&=\rho_d \beta_l^2\left(2\sum\limits_{\substack{i=0}}^{N-1}{\hat \gamma''_{lll}(i)}-1\right)+\rho_d\beta_l\sum\limits_{\substack{j=1\\j\neq l}}^{K}{\beta_j\sum\limits_{\substack{i=0}}^{N-1}\hat \gamma''_{ljl}(i)}\\
\nonumber
IUI&=\rho_d \sum\limits_{\substack{k=1\\k\neq l}}^{K}\beta_k^2\sum\limits_{\substack{i=0}}^{N-1}{2\hat \gamma''_{lkk}(i)}+\rho_d\sum\limits_{\substack{k=1\\k\neq l}}^K\sum\limits_{\substack{j=1\\j\neq k}}^{K}{\beta_{k}\beta_{j}\sum\limits_{\substack{i=0}}^{N-1}\hat \gamma''_{ljk}(i)}\\
\nonumber
noise&=\frac{\rho_d}{\rho_p}{u_{l0}}\sum\limits_{\substack{k=1}}^{K}{\beta_k}+\left(\sum_{k=1}^{K}{\beta_{k}\lambda''_{lk}}+\frac{1}{\rho_p}\right){v_{l0}}
\end{flalign}
\normalsize
where ${u_{l(p-q)}}=[\boldsymbol{W_{ll}}E[\boldsymbol{\hat G_{k}}\boldsymbol{\hat G}^H_{\boldsymbol{k}}]\boldsymbol{W}^H_{\boldsymbol{ll}}](p,q)$ and ${v_{l(p-q)}}=[\boldsymbol{W_{ll}}\boldsymbol{\Sigma_{n^{os}}}\boldsymbol{W}^H_{\boldsymbol{ll}}](p,q)$. Also,
\small
\begin{flalign}
\hat \gamma''_{ljk}(p-q)=\int_{-\infty}^{\infty}\int_{-\infty}^{\infty}{\left(\lambda_{lj}^s \boldsymbol{\hat{G}_{lk}}(p.q)\right)^2f(\tau_j)f(\tau_k)d\tau_jd\tau_k}
\end{flalign}
\begin{flalign}
\lambda''_{lk}=\int_{-\infty}^{\infty}{\left(\lambda_{lk}^s \right)^2f(\tau_k)d\tau_k}
\end{flalign}
\normalsize
where $\boldsymbol{\hat{G}_{lk}}={\boldsymbol{W}_{ll}\boldsymbol{\hat G_{k}}}$ and $\boldsymbol{\hat G_{k}}=[(\boldsymbol{G^1_{k}})^T,\cdots,(\boldsymbol{G^K_{k}})^T]^T$. Assuming
the same distribution for all the time delays, the receive antenna index is discarded.   
\end{theorem}
\begin{IEEEproof}
The proof is presented in Appendix \ref{appendix_4}.
\end{IEEEproof}
By using the MRC-ZF receiver which exploits the statistics of unknown time delays, the effect of averaged ISI and IUI is vanished. If the number of receive antennas goes to infinity, the achievable rate goes to infinity. Replacing $\rho_d=\frac{E_d}{\sqrt{M}}$ in Eq. (\ref{mrc_zf2}) and letting $M$ go to infinity, we will have:
\small
\begin{flalign}
R_{l,ip}^{mrc-zf}\rightarrow\kappa\log_{2}{\left(1+\frac{E_d\beta_l}{{v_{l0}}}\right)},\textmd{as} \ M\rightarrow \infty, \rho_d=\frac{E_d}{\sqrt{M}}
\end{flalign}
\normalsize
Hence, even in the presence of unknown time delays, the power scaling law is held by using the MRC-ZF receiver. The value of ${v_{l0}}$ is calculated based on the pulse shape, the time delay distribution and the sampling origin. As mentioned before, the loss of ${v_{l0}}$ which is because of the noise enhancement by ZF can be mitigated by using other cancellation methods like MMSE and SIC. 

Note that in the multi-cell scenario, not only the users within the corresponding cell but also the users from the adjacent cells will cause interference. In the perfect CSI scenario, the additional interference from the users in other cells is $\rho_d\sum\limits_{i=-I}^{I}{E[g_i^2]}\sum\limits_{\substack{k=1 }}^{K'}{\beta'_k}$, where $K'$ is the number of interfering users outside the desired cell and $\beta'_k$ is the path-loss of the corresponding user. By performing the same ZF cancellation, we can remove the effect of the ISI as $M\rightarrow \infty$ and restore the benefits of the massive MIMO systems. For the imperfect CSI scenario, the additional interference term is defined similar to the IUI term in Theorem \ref{22}. If the same pilot sequence is used for two interfering cells, namely 1 and 2, matrix $\boldsymbol{\Gamma_l}$ in Eq. (\ref{71}) which only depends on sampling times, pilot sequences, the pulse shape and delay distributions will be the same for both within the cell and outside the cell users, i.e. $\boldsymbol{\Gamma^1_l}=\boldsymbol{\Gamma^2_l}$. Therefore, by using the MRC-ZF receiver, we can only cancel the effect of the  ISI and the intra-cell interference, however, the effect of the inter-cell interference caused by the pilot reuse will remain. To address this issue, one can add distinct intentional delays to the transmissions of each cell, in such a way that $\boldsymbol{\Gamma^1_l} \neq \boldsymbol{\Gamma^2_l}$. Then, by utilizing the same concept of oversampling and ZF, as described in Section \ref{pool}, the effect of the inter-cell interference can also be mitigated.

\section{Simulation Results}\label{sec-simu}
\begin{figure}[!h]
\centering
\includegraphics[width=6in]{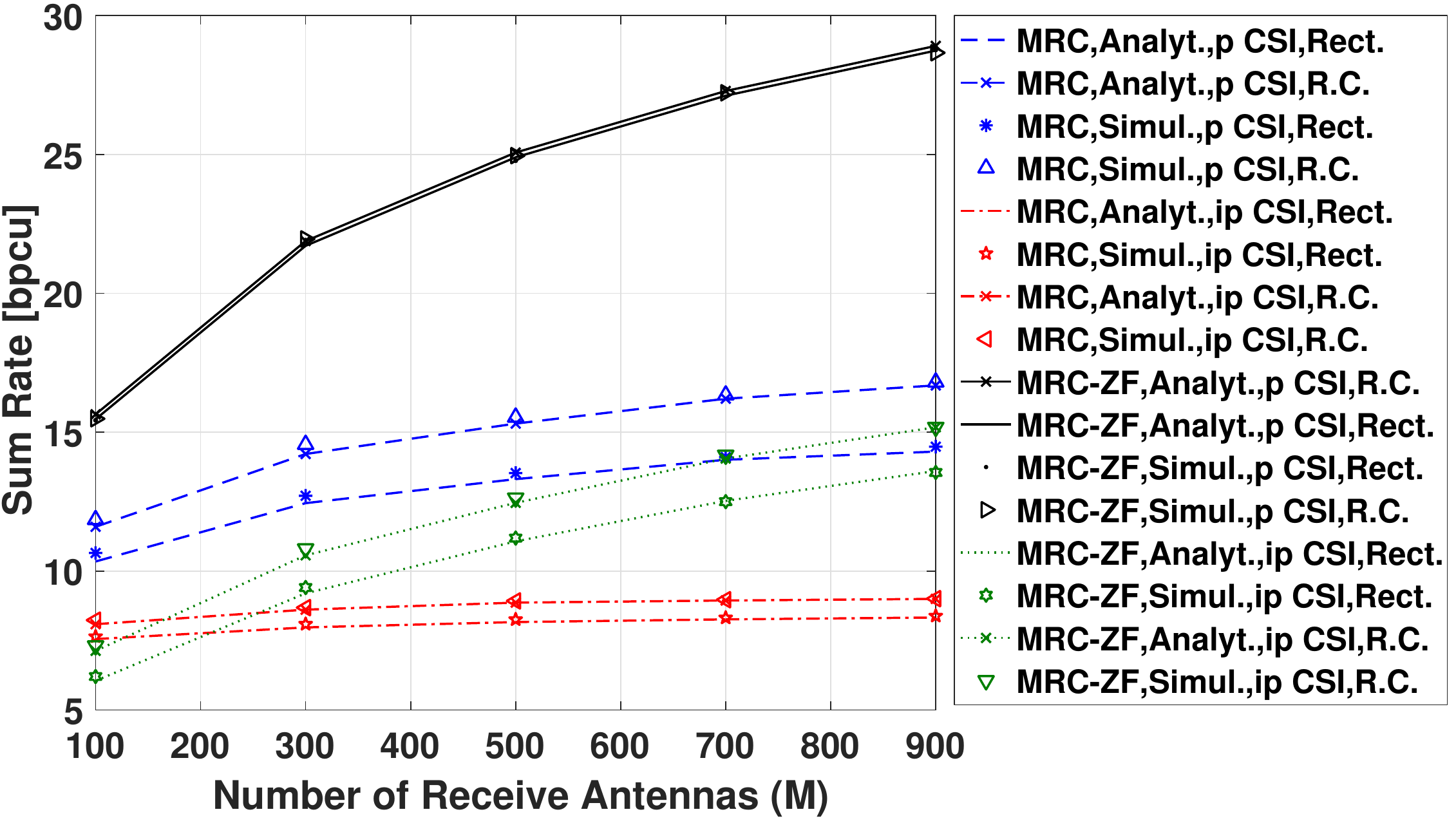}
\caption{Performance of the MRC and MRC-ZF  receivers with respect to the number of receive antennas, for 5 users each of them using 20 dB transmit power}
\label{all}
\end{figure}

In this section, simulation results are presented to verify our theoretical analysis. In the simulations, time delays follow the distribution mentioned in Eq. (\ref{delays}), noise samples and fading coefficients are also distributed as $CN(0,1)$. The large-scale
channel fading is modeled as $\beta_k=z_k/(r_k/r_h)^v$, where $z_k$
is a log-normal random variable with standard deviation of $\sigma$, $v$ is the path-loss exponent, and $r_k$ is the distance between the $k$th user and the BS which varies in the interval of $[r_h,R]$. We have assumed that $v=1.8$, $\sigma=8$, $r_h=100$ and $R=1000$. The large scale fading is assumed to be fixed and $10,000$ different realizations of time delays, fading coefficients and noise samples are simulated. In Fig. \ref{all}, the performance of the MRC and MRC-ZF receivers with perfect CSI and imperfect CSI is presented by theoretical approximation in Theorems \ref{11}, \ref{22}, \ref{33} and \ref{44} and via simulation. The sum rate for 5 users are plotted with respect to the number of receive antennas. The results include rectangular (Rect.) pulse shape and raised cosine (R.C.) pulse shape with roll-off factor of $\beta=0.5$ truncated at $3$ side lobes. Our theoretical approximation and simulation results match. It also shows that, unknown time delays limit the performance of MRC receivers, and by increasing $M$, the sum rate is saturated. However, the performance of the MRC-ZF receiver is not saturated and by increasing $M$, the sum rate increases. 

\begin{figure*}[!t]
\centering
\subfloat[perfect CSI]{\includegraphics[width=3in]{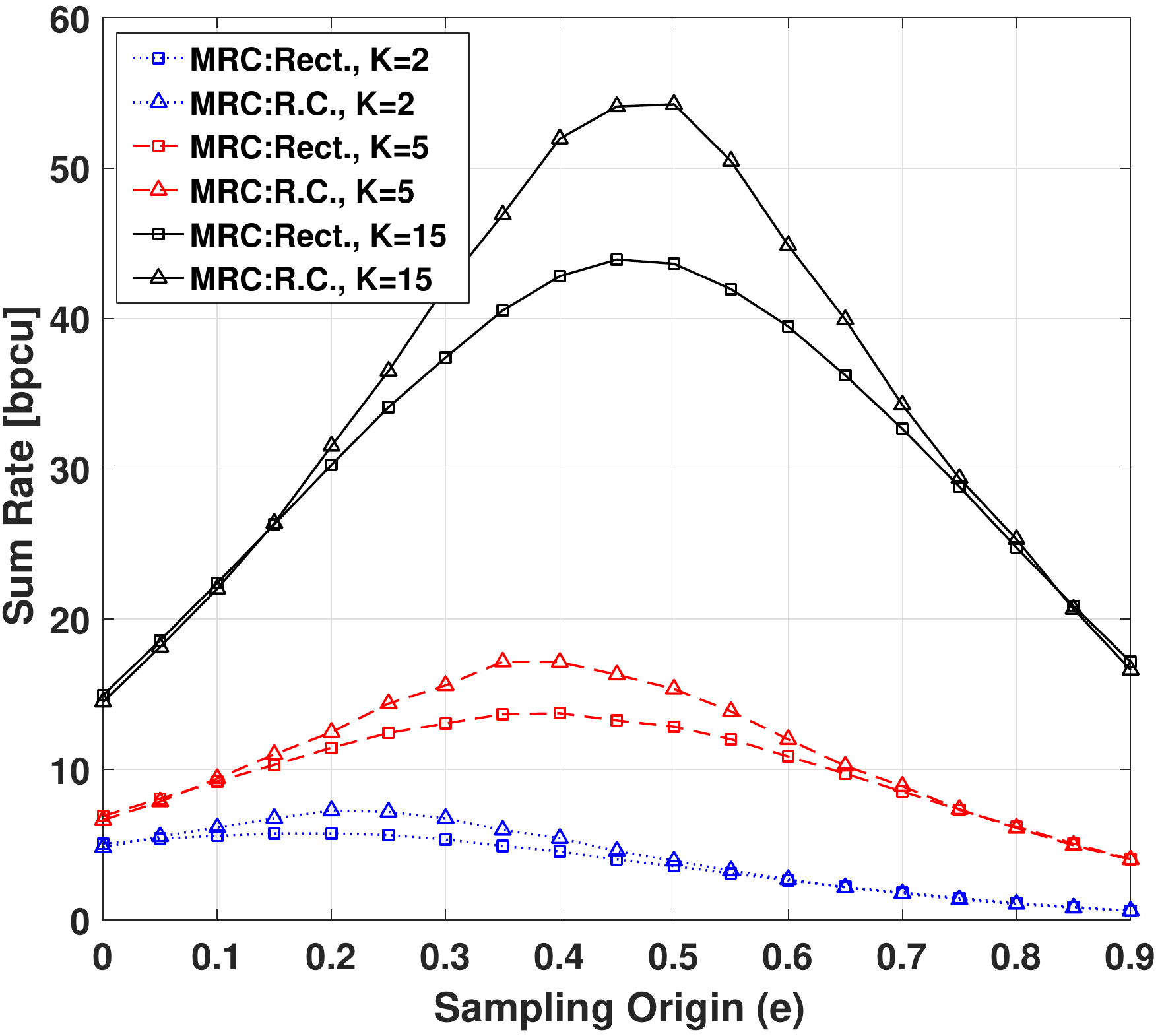}
\label{e_p}}
\hfil
\subfloat[imperfect CSI]{\includegraphics[width=3in]{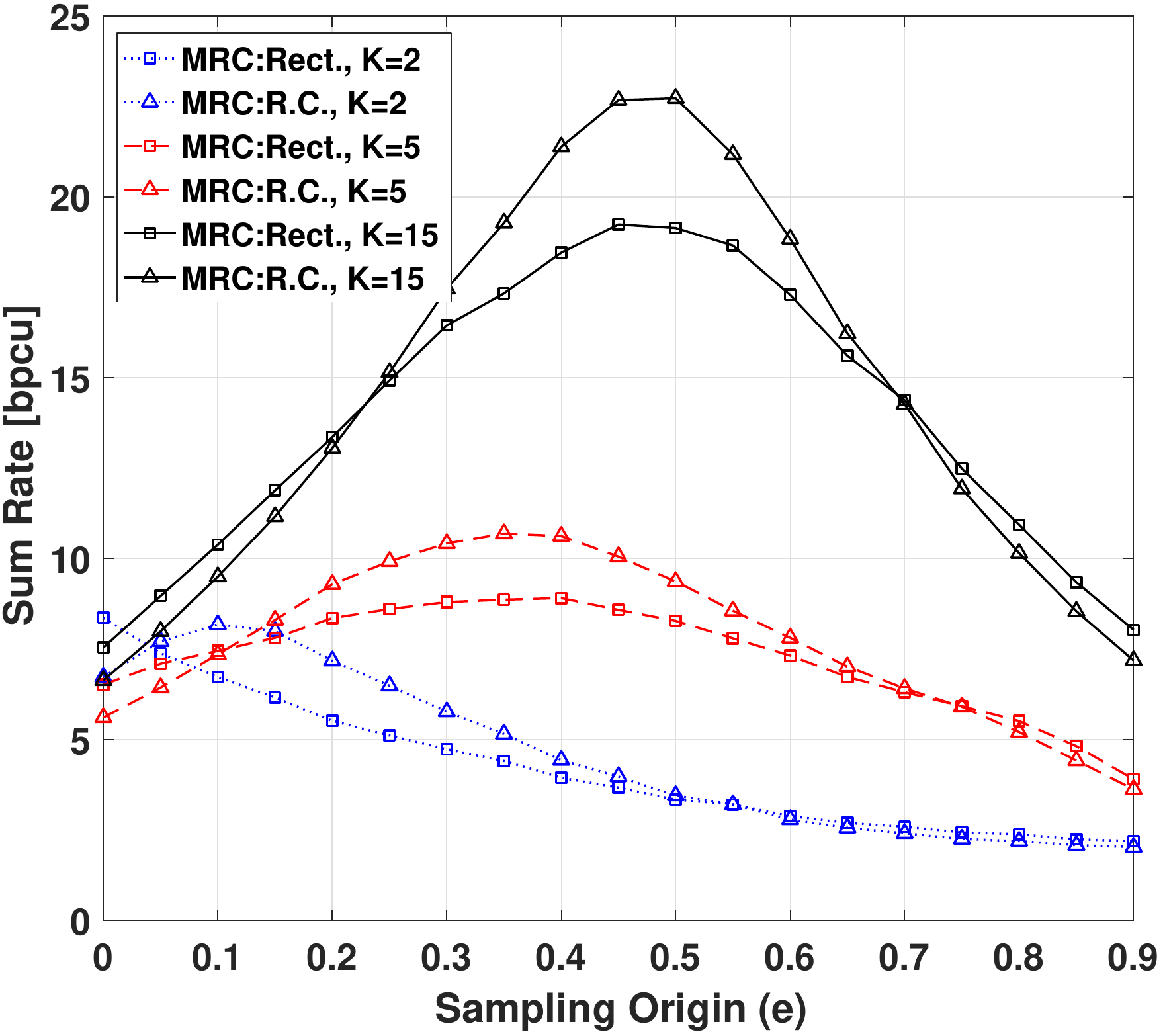}
\label{e_ip}}
\caption{Asymptotic performance of the MRC receiver with respect to sampling origin for different number of users, using perfect/imperfect  CSI and $\rho_d=20$ dB}
\end{figure*}

Fig. \subref*{e_p} shows the asymptotic performance, i.e. $M\rightarrow \infty $, with respect to the sampling origin, i.e., $e$. It includes the result for different number of users, $K=[2,5,15]$, and as the number of users increases, the optimal value of $e$ tends to half, which verifies the results in Table \ref{table1}. This is in line with the fact that for small number of users the distribution in Eq. (\ref{delays}) is closer to a delta function whose optimal sampling origin is $e=0$, and by increasing the number of users, the distribution tends to uniform whose best sampling origin is half.

\begin{figure*}[!t]
\centering
\subfloat[perfect CSI, $\rho_d=\frac{E_d}{M}$ and $M\rightarrow \infty$]{\includegraphics[width=3in]{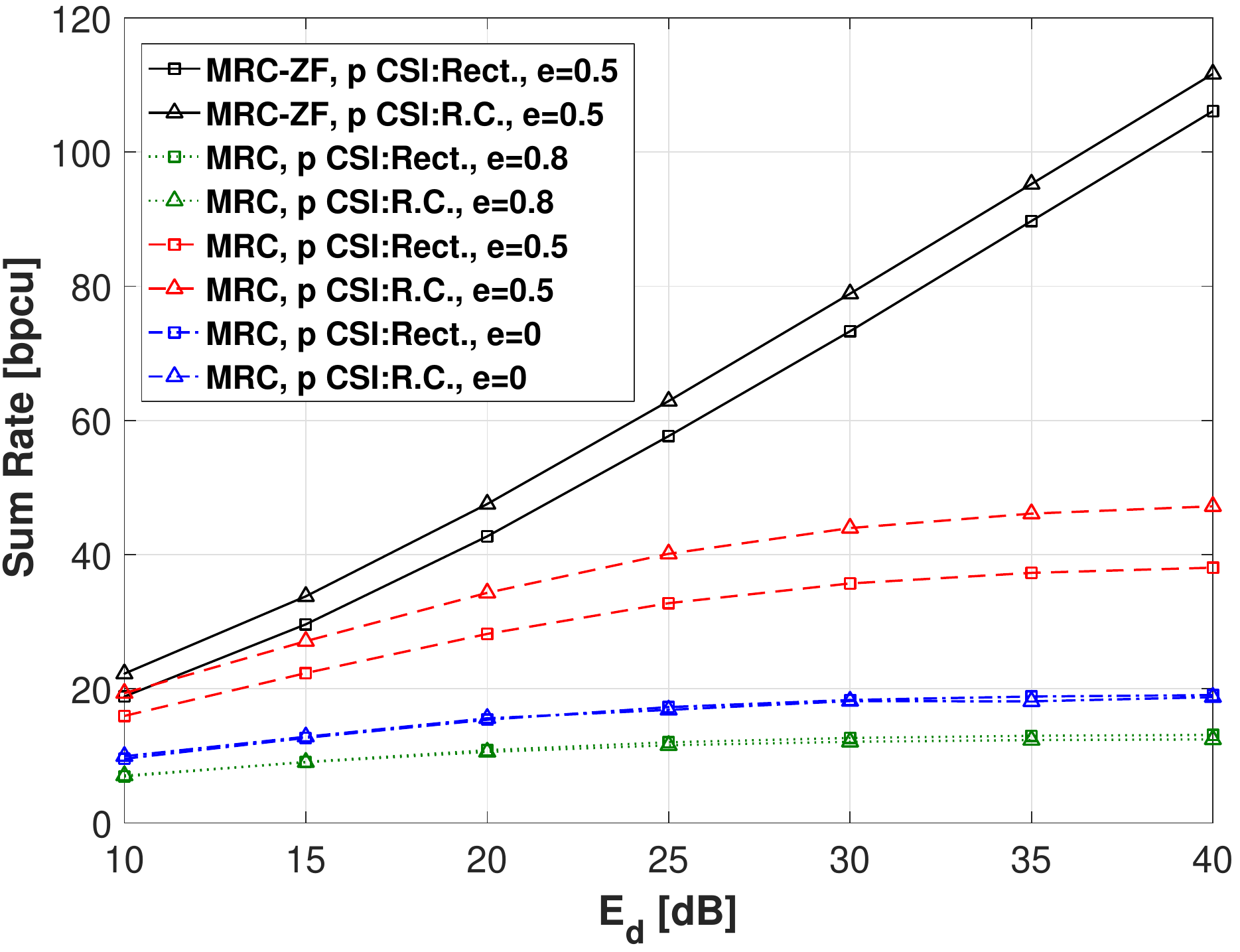}
\label{compare_p}}
\hfil
\subfloat[imperfect CSI, $\rho_d=\frac{E_d}{\sqrt{M}}$ and $M\rightarrow \infty$]{\includegraphics[width=3in]{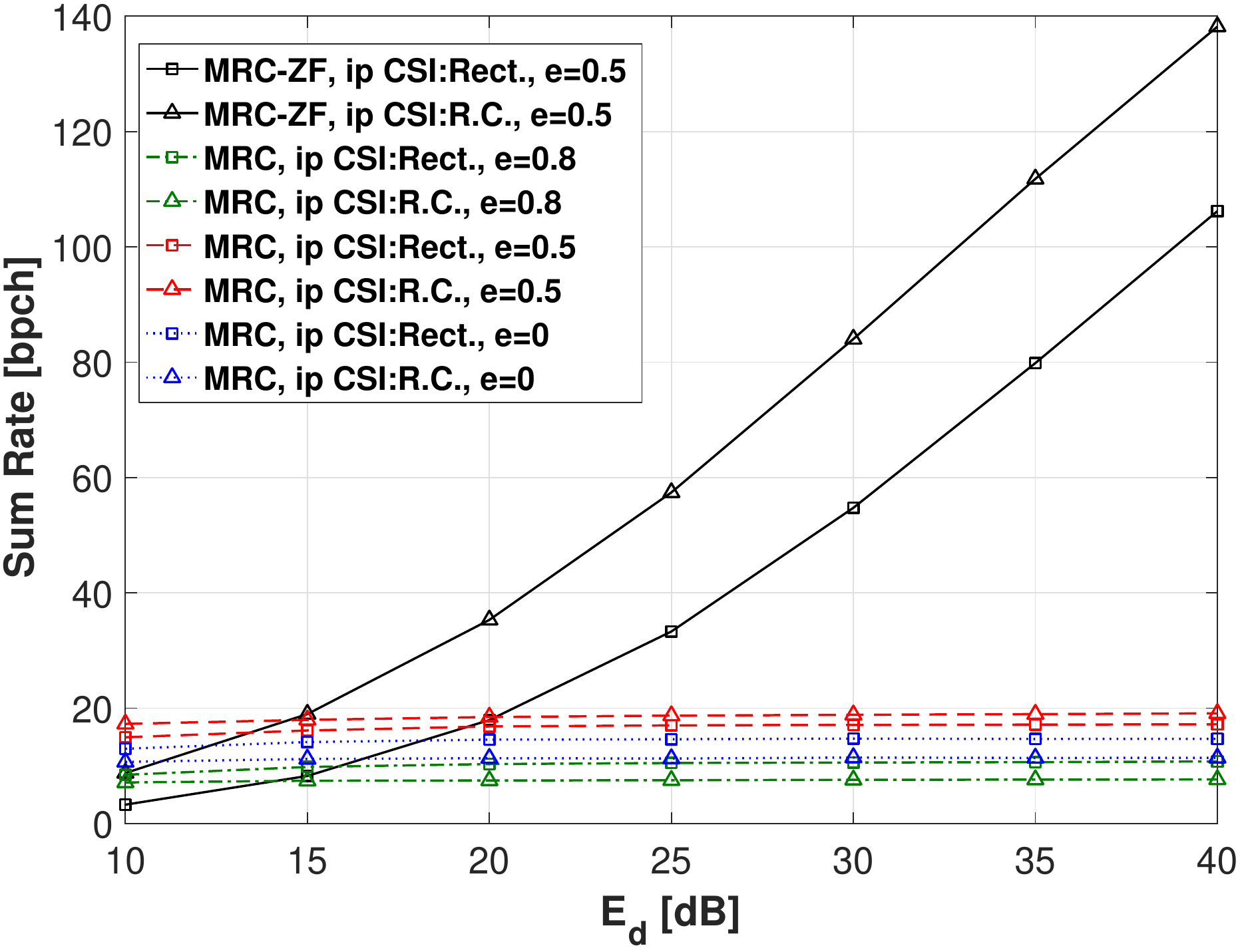}
\label{compare_ip}}
\caption{Comparison of the MRC and the MRC-ZF receiver for 10 users}
\end{figure*}

In Fig. \subref*{e_ip}, a similar analysis is presented for imperfect CSI case. Again, we observe that, changing $e$ can change the achievable rate, significantly, and by increasing the number of users, its optimal value tends to half. 
In Fig. \subref*{compare_p}, the performance of the MRC and the MRC-ZF receivers are presented while $\rho_d=\frac{E_d}{M}$ and $M\rightarrow \infty$. Increasing $E_d$ does not increase the sum rate achieved by the MRC receiver. On the other hand, by using the MRC-ZF receiver, the achievable sum rate increases which shows that the power-scaling law can be achieved.

In Fig. \subref*{compare_ip}, a similar analysis is presented for imperfect CSI with a power scale of  $\frac{1}{\sqrt{M}}$. Again, it verifies our theoretical results that the MRC receiver is unable to hold the power scaling law when there is time delay between received signals. However, by implementing the MRC-ZF, a power scale of $\frac{1}{\sqrt{M}}$ is achieved.

\section{Conclusion}\label{sec-con}
In this work, we obtained the general formula for the achievable rate by the MRC receiver when unknown timing mismatch exists. We showed that unknown timing mismatch degrades the performance substantially. In other words, in the presence of unknown timing mismatch, the achievable rate by each user is limited when the number of receive antennas goes to infinity. To address this challenge, we introduced two receiver design methods, one when the perfect CSI is available and one when CSI is estimated by orthogonal pilot sequences, which restore the benefits of massive MIMO. We proved that these introduced receivers follow the power scaling law, i.e., the single-user performance with no interference can be achieved by using an arbitrary small transmit power. In our proposed receiver designs, we used ZF to cancel the effect of averaged ISI and IUI, however, other methods like MMSE or SIC can also be used. We also used oversampling besides ZF to cancel the effect of contamination in the channel estimations caused by timing mismatch. At the end, we presented simulation results which confirmed our analysis. 

\appendices
\section{}\label{appendix_main}
We analyzed four different receiver structures, including: MRC with perfect and imperfect CSI and MRC-ZF with perfect and imperfect CSI. The output sample of either of these receivers for detection of the $a$th symbol of the $l$th user can be written in a general framework as:
\begin{flalign}
\boldsymbol{{y}_{l,p/ip}^{mrc/mrc-zf}}(a)=\sqrt{\rho_d}\sum_{k=1}^{K}\sum_{n=1}^{N}{\boldsymbol{T_{lk,p/ip}^{mrc/mrc-zf}}(a,n)\boldsymbol{b_k}(n)+\boldsymbol{n_{l,p/ip}^{mrc/mrc-zf}}(a)}
\label{far}
\end{flalign}
Discarding the subscripts for different receivers and based on the assumption that the coefficients are known perfectly at the receiver, the corresponding achievable rate can be calculated as follows\cite{ngo2013energy}:
\small
\begin{flalign}\label{asli}
R_{l}(a)=E\left[\log_2\left(1+\frac{\rho_d(\boldsymbol{T_{ll}}(a,a))^2}{\rho_d\sum\sum_{\substack{k,n\\(k,n) \neq (l,a)}}{(\boldsymbol{T_{lk}}(a,n))^2}+\sigma^2_{n_l(a)}}\right)\right]
\end{flalign}
\normalsize
Because each of the coefficients is the arithmetic mean of $M$ uncorrelated random variables, as $M$ increases, their randomness decreases and they approach to their expected value. As a result, the above expression can be approximated in the Massive MIMO context as follows: \cite{rate_massiveMIMO,krishnan2016linear}
\small
\begin{flalign}\label{asli2}
 R^{approx}_{l}(a)=\log_2\left(1+\frac{\rho_dE\left[(\boldsymbol{T_{ll}}(a,a))^2\right]}{\rho_d\sum\sum_{\substack{k,n\\(k,n) \neq (l,a)}}{E\left[(\boldsymbol{T_{lk}}(a,n))^2\right]}+\sigma^2_{n_l(a)}}\right)
\end{flalign}
\normalsize
The upper-bound for the approximation error is also presented in Appendix \ref{appendix_approx}. However, due to the existence of unknown time delays, the assumption of perfect knowledge of coefficients at the receiver is not valid anymore. Because the expected value of the coefficients is known by the receiver, in order to find the achievable rates, we rewrite Eq. (\ref{far}), as follows (the subscripts are discarded):
\begin{flalign}
\nonumber
\boldsymbol{{y}_{l}}(a)&=\sqrt{\rho_d}\sum_{k=1}^{K}\sum_{n=1}^{N}{E[\boldsymbol{T_{lk}}(a,n)]\boldsymbol{b_k}(n)}+\sqrt{\rho_d}\sum_{k=1}^{K}\sum_{n=1}^{N}{\left(\boldsymbol{T_{lk}}(a,n)-E[\boldsymbol{T_{lk}}(a,n)]\right)\boldsymbol{b_k}(n)}+\boldsymbol{n_{l}}(a)\\
&=\sqrt{\rho_d}\sum_{k=1}^{K}\sum_{n=1}^{N}{E[\boldsymbol{T_{lk}}(a,n)]\boldsymbol{b_k}(n)}+\boldsymbol{\tilde n_{l}}(a)
\label{far2}
\end{flalign}
In this new system model, all the coefficients are known by the receiver and the effective noise is denoted as $\boldsymbol{\tilde {n_l}}$. It is mentioned in \cite{pitarokoilis2012optimality} that the achievable information rate
for the system model in Eq. (\ref{far2}) is given by considering the worst
case uncorrelated additive noise having the same variance as $\boldsymbol{\tilde n_{l}}(a)$. Variance of $\boldsymbol{\tilde n_{l}}(a)$ can be easily calculated as $\sigma^2_{\tilde n_l(a)}=\rho_d\sum_{k=1}^{K}\sum_{n=1}^{N}{Var[\boldsymbol{T_{lk}}(a,n)]}+\sigma^2_{n_l(a)}$. As a result, the achievable rate can be written as:
\small
\begin{flalign}
\tilde R_l(a)=E\left[\log_2{\left(1+\frac{\rho_dE^2[\boldsymbol{T_{ll}}(a,a)]}{\rho_d\sum\sum_{\substack{k,n\\(k,n) \neq (l,a)}}{E^2\left[\boldsymbol{T_{lk}}(a,n)\right]}+\rho_d\sum_{k=1}^{K}\sum_{n=1}^{N}{Var[\boldsymbol{T_{lk}}(a,n)]}+\sigma^2_{n_l(a)}}\right)}\right]
\end{flalign}
\normalsize
The expectation can be discarded because the randomness of the coefficients is combined with the noise and no randomness is left except the noise realizations where Gaussian uncorrelated additive noise is considered. Then,
\small
\begin{flalign}
\tilde R_l(a)=\log_2{\left(1+\frac{\rho_dE^2[\boldsymbol{T_{ll}}(a,a)]}{\rho_d\sum\sum_{\substack{k,n\\(k,n) \neq (l,a)}}{E^2\left[\boldsymbol{T_{lk}}(a,n)\right]}+\rho_d\sum_{k=1}^{K}\sum_{n=1}^{N}{Var[\boldsymbol{T_{lk}}(a,n)]}+\sigma^2_{n_l(a)}}\right)}
\end{flalign}
\normalsize
By using the fact that $Var[x]=E[x^2]-E^2[x]$, the formula for the achievable rate can be further simplified to:
\small
\begin{flalign}
\tilde R_l(a)=\log_2{\left(1+\frac{\rho_dE^2[\boldsymbol{T_{ll}}(a,a)]}{\rho_d\sum_{k=1}^K\sum_{n=1}^N{E\left[(\boldsymbol{T_{lk}}(a,n))^2\right]}-\rho_dE^2[\boldsymbol{T_{ll}}(a,a)]+\sigma^2_{n_l(a)}}\right)}
\label{far3}
\end{flalign}
\normalsize
Note that the only difference with the approximation in Eq. \ref{asli2} is that $VAR[\boldsymbol{T_{ll}}(a,a)]$ is subtracted from the nominator and is added to the denominator. For finding the achievable rates for different receivers, we just need to calculate the values of $E[(\boldsymbol{T_{lk}}(a,n))^2]$ and the variance of effective noise vector for different receivers. Note that due to the structure of the system, the achievable rate for different
symbols of the frame except the $I$-boundary ones (negligible with respect to the frame length) is the same, thus the index of $a$ can be
discarded.

\section{}\label{appendix_1}
Using the results presented in Appendix \ref{appendix_main}, we show the step by step derivation for the MRC receiver with perfect CSI. The corresponding formulas for other scenarios can be derived similarly. For the MRC receiver with perfect CSI the effective channel matrix is denoted as  $\boldsymbol{T^{mrc}_{lk,p}}=\frac{1}{M}\sum_{m=1}^{M}\sqrt{\beta_l \beta_k}h^*_{lm}h_{km}\boldsymbol{G_{km}}$ and the values of $E[|\boldsymbol{T^{mrc}_{lk,p}}(a,n)|^2]$ are calculated in the next lemma.
\begin{lemma}\label{hey1}
The expected value of $|\boldsymbol{T_{lk,p}^{mrc}}(a,n)|^2$ can be calculated as follows:
\begin{flalign}
E\left[|\boldsymbol{T_{lk,p}^{mrc}}(a,n)|^2\right]=\left\{\begin{matrix}
\frac{\beta^2_l}{M}\left(2E[g_{a-n}^2]+(M-1)E^2[g_{a-n}]\right)& k=l\\ 
\frac{\beta_l\beta_k}{M}(E[g_{a-n}^2]) & k\neq l
\end{matrix}\right.
\end{flalign}
where 
\begin{flalign}\label{zolfaye_leili}
{E}[g^r_i]=\int_{-\infty}^{\infty}{g^r(e+T+iT_s-\tau)f(\tau)d\tau}
\end{flalign}
\end{lemma}

\begin{IEEEproof}
Case $k=l$:
\small
\begin{flalign}
\nonumber
&E\left[|\boldsymbol{T_{ll,p}^{mrc}}(a,n)|^2\right]=\frac{1}{M^2}E\left[ \left(\sum_{m=1}^{M}{\beta_l|h_{lm}|^2\boldsymbol{G_{lm}}(a,n)}\right) \left(\sum_{m=1}^{M}{\beta_l|h_{lm}|^2\boldsymbol{G_{lm}}(a,n)}\right) \right]\\
\nonumber
&=\frac{\beta^2_l}{M^2}\left(\sum_{m=1}^ME\left[|h_{lm}|^4\boldsymbol{G}^2_{\boldsymbol{lm}}(a,n)\right]+\sum_{m_1=1}^M\sum_{\substack{m_2=1\\m_2 \neq m_1}}^ME\left[|h_{lm_1}|^2\boldsymbol{G}_{\boldsymbol{lm_1}}(a,n)|h_{lm_2}|^2\boldsymbol{G}_{\boldsymbol{lm_2}}(a,n)\right]\right)\\
&=\frac{\beta^2_l}{M^2}\left(2\sum_{m=1}^ME\left[\boldsymbol{G}^2_{\boldsymbol{lm}}(a,n)\right]+\sum_{m_1=1}^M\sum_{\substack{m_2=1\\m_2 \neq m_1}}^ME\left[\boldsymbol{G}_{\boldsymbol{lm_1}}(a,n)\boldsymbol{G}_{\boldsymbol{lm_2}}(a,n)\right]\right)\label{talkh}
\end{flalign}
\normalsize
where the expectation of the elements of matrices $\boldsymbol{G}_{\boldsymbol{lm}}$ is taken over different realizations in time. These expectations only depend on the pulse shape and the joint distribution of time delays which are known by the receiver. For example, $E\left[\boldsymbol{G}_{\boldsymbol{lm_1}}(a,n)\boldsymbol{G}_{\boldsymbol{lm_2}}(a,n)\right]$ can be expressed as:
\small
\begin{flalign}
E\left[\boldsymbol{G}_{\boldsymbol{lm_1}}(a,n)\boldsymbol{G}_{\boldsymbol{lm_2}}(a,n)\right]=\int_{-\infty}^{\infty}\int_{-\infty}^{\infty}{g(t'-\tau_{1})g(t'-\tau_{2})f_{\tau_{lm_1},\tau_{lm_2}}(\tau_1,\tau_2)d\tau_1d\tau_2}
\end{flalign}
\normalsize
where $t'=e+T+(a-n)T_s$. Note that based on the system characteristics, the time delays can follow any given correlated distribution or even they might be the same which is more suited for collocated receive antennas , i.e., $f_{\tau_{lm_1},\tau_{lm_2}}(\tau_1,\tau_2)=f(\tau_1)\delta(\tau_1-\tau_2)$, or independent for distributed receive antennas, i.e., $f_{\tau_{lm_1},\tau_{lm_2}}(\tau_1,\tau_2)=f_{\tau_{lm_1}}(\tau_1)f_{\tau_{lm_2}}(\tau_2)$. Although no assumption is required for the joint distribution of delays, we consider the distributed scenario and assume that all the time delays are independent and follow the same distribution, i.e., $f_{\tau_{lm}}(\tau)=f(\tau)$. Therefore, we will have:
\small
\begin{flalign}
\nonumber
E\left[\boldsymbol{G}_{\boldsymbol{lm_1}}(a,n)\boldsymbol{G}_{\boldsymbol{lm_2}}(a,n)\right]=\left[\int_{-\infty}^{\infty}{g(t'-\tau)f(\tau)d\tau}\right]^2,\ E\left[\boldsymbol{G}^2_{\boldsymbol{lm}}(a,n)\right]=\int_{-\infty}^{\infty}{g^2(t'-\tau)f(\tau)d\tau}
\end{flalign}
\normalsize
Note that, $E\left[\boldsymbol{G}_{\boldsymbol{lm}}(a,n)\right]$ depends on $a$ and $n$ thorough their difference, therefore we denote it as $E[g_{a-n}]$. As a result, Eq. (\ref{talkh}) can be presented as:
\begin{flalign}
\nonumber
&E\left[|\boldsymbol{T_{ll,p}^{mrc}}(a,n)|^2\right]=\frac{\beta^2_l}{M^2}\left(2ME\left[g^2_{a-n}\right]+M(M-1)E^2\left[g_{a-n}\right]\right)
\end{flalign}
Case $k\neq l$: By taking similar steps, we can show that:
\small
\begin{flalign}
\nonumber
E\left[|\boldsymbol{T_{lk,p}^{mrc}}(a,n)|^2\right]&=\frac{1}{M^2}E\left[ \left(\sum_{m=1}^{M}{\sqrt{\beta_l\beta_k}h_{lm}^*h_{km}\boldsymbol{G_{km}}(a,n)}\right) \left(\sum_{m=1}^{M}{\sqrt{\beta_l\beta_k}h_{lm}h_{km}^*\boldsymbol{G_{km}}(a,n)}\right) \right]\\
\nonumber
&=\frac{\beta_l\beta_k}{M^2}\left(\sum_{m=1}^ME\left[|h_{lm}|^2|h_{km}|^2\boldsymbol{G}^2_{\boldsymbol{km}}(a,n)\right]\right)\\
\nonumber
&=\frac{\beta_l\beta_k}{M^2}\left(\sum_{m=1}^ME\left[\boldsymbol{G}^2_{\boldsymbol{km}}(a,n)\right]\right)\\
&=\frac{\beta_l\beta_k}{M^2}(ME[g_{a-n}^2])
\end{flalign}
\normalsize
which completes the derivation of expectations.
\end{IEEEproof}
Covariance matrix of the effective noise vector is also calculated as:
\small
\begin{flalign}
\nonumber
COV[\boldsymbol{n_{l,p}^{mrc}}]&=E\left[\boldsymbol{n_{l,p}^{mrc}}\boldsymbol{n_{l,p}^{mrc}}^H\right]\\
&=E\left[\left(\frac{\sqrt{\beta_l}}{M}\sum_{m=1}^M h^*_{lm} \boldsymbol{n_m}\right)\left(\frac{\sqrt{\beta_l}}{M}\sum_{m=1}^M h_{lm} \boldsymbol{n}^H_{\boldsymbol{m}}\right)\right]\\
&=\frac{\beta_l}{M^2}\sum_{m=1}^ME\left[|h_{lm}|^2\right]E\left[\boldsymbol{n_m}\boldsymbol{n}^H_{\boldsymbol{m}}\right]\\
&=\frac{\beta_l}{M}\boldsymbol{I_N}
\end{flalign}
\normalsize
By inserting the expected values of $|\boldsymbol{T_{lk,p}^{mrc}}(a,n)|^2$ and also the variance of the effective noise vector into Eq. (\ref{far3}), we have:
\small
\begin{flalign}
\tilde R_{l,p}^{mrc}=\log_{2}{\left(1+\frac{\rho_d \beta^2_lE^2[g_0]}{\rho_d\sum\limits_{i=-I}^{I}{E[g_i^2]}\sum\limits_{\substack{k=1}}^{K}{\beta_k}+\rho_d \beta_l\sum\limits_{\substack{i=-I }}^{I}(E[g_i^2]+(M\bar \delta[i]-1)E^2[g_i])+1}\right)}
\end{flalign}
\normalsize
where $\bar\delta[i]=1-\delta[i]=\left\{\begin{matrix}
1 \ \ i\neq 0\\ 
0 \ \ i=0
\end{matrix}\right.$. Note that $E[g_i]$ is nonzero only for $-I\leq i\leq I$.  

\section{}\label{appendix_2}
By taking similar steps as Appendix \ref{appendix_1}, we can derive the expectations of the elements of the effective matrix presented in Eq. (\ref{effective_matrix}), as follows:
\small
\begin{flalign}
\nonumber
E\left[|\boldsymbol{T_{lk,ip}^{mrc}}(a,n)|^2\right]=\frac{1}{M^2}\left\{ \vphantom{ \sum_{\substack{j=1\\j \neq l}}^{K}}\beta^2_k\left(2M\gamma''_{lkk}(a-n)+M(M-1)(\gamma'_{lkk}(a-n))^2\right) + \sum_{\substack{j=1\\j \neq k}}^{K}\beta_j\beta_k(M\gamma''_{ljk}(a-n)) \right\}
\end{flalign}
\normalsize
where $\gamma''_{ljk}(a-n)=E[\gamma^2_{ljkm}(a-n)]$ and $\gamma'_{ljk}(a-n)=E[\gamma_{ljkm}(a-n)]$ are the expectations over the time delay distributions. Assuming
the same distribution for all the time delays, the receive antenna index is discarded. The covariance matrix of the effective noise vector presented in Eq. (\ref{noise_ip}) can be calculated similarly as:
\small
\begin{flalign}
COV[\boldsymbol{n_{l,ip}^{mrc}}]=\left(\frac{\rho_d}{M\rho_p}\sum\limits_{\substack{k=1}}^{K}{\beta_k\sum\limits_{\substack{i=-w}}^{w}E[g^2_i]}+\frac{1}{M}\left(\sum_{j=1}^{K}\beta_j\lambda''_{lj}+\frac{1}{\rho_p}\right)\right)\boldsymbol{I_N}
\end{flalign}
\normalsize
where $\lambda''_{lj}=E[\lambda^2_{ljm}]$ is the average of $\lambda^2_{ljm}$ over the time delay distributions. By inserting these values into Eq. (\ref{far3}), the proof will be complete.

\section{}\label{appendix_3}
The expected values of  $|\boldsymbol{T_{lk,p}^{mrc-zf}}(a,n)|^2$ defined in Eq. (\ref{sys_mrczf}) can be calculated similar to Appendix \ref{appendix_1}:
\small
\begin{flalign}
E\left[|\boldsymbol{T_{lk,p}^{mrc-zf}}(a,n)|^2\right]=\left\{\begin{matrix}
\frac{\beta^2_l}{M^2}(2M\xi''_{a-n}+M(M-1)\delta(a-n))& k=l\\ 
\frac{\beta_l\beta_k}{M^2}(M\xi''_{a-n}) & k\neq l
\end{matrix}\right.
\end{flalign}
\normalsize
where $\xi''_{a-n}=E\left[\boldsymbol{G'}^2_{\boldsymbol{}}(a,n)\right]$ and $\boldsymbol{G'}=\boldsymbol{ZG}$. Covariance of the effective noise vector in Eq. (\ref{sys_mrczf}) is also calculated as:
\small
\begin{flalign}
COV[\boldsymbol{n_{l,p}^{mrc-zf}}]=E\left[\boldsymbol{n_{l,p}^{mrc-zf}}\boldsymbol{n_{l,p}^{mrc-zf}}^H\right]=E\left[\boldsymbol{Zn_{l,p}^{mrc}}\boldsymbol{n_{l,p}^{mrc}}^H\boldsymbol{Z}^H\right]
=\frac{\beta_l}{M}\boldsymbol{Z}\boldsymbol{Z}^H
\end{flalign}
\normalsize
By inserting these values into Eq. (\ref{far3}), we can conclude the proof.

\section{}\label{appendix_4}
Similar to Appendix \ref{appendix_1}, the expected values of $|\boldsymbol{T_{lk,ip}^{mrc-zf}}(a,n)|^2$  and the covariance matrix of the effective noise vector defined in Eq. (\ref{milk_choc}) can be calculated, respectively, as:
\small
\begin{flalign}
\nonumber
E\left\{|\boldsymbol{T_{lk,ip}^{mrc-zf}}(a,n)|^2\right\}=
\frac{1}{M^2}\left\{ \vphantom{ \sum_{\substack{j=1\\j \neq l}}^{K}}\beta^2_k\left(2M\hat \gamma''_{lkk}(a-n)+M(M-1)\delta(l-k)\boldsymbol{I_N}(a,n)\right) + \sum_{\substack{j=1\\j \neq k}}^{K}\beta_j\beta_k(M\hat \gamma''_{ljk}(a-n)) \right\}
\end{flalign}
\begin{flalign}
COV[\boldsymbol{n_{l,ip}^{mrc-zf}}]=\frac{\rho_d}{M\rho_p}\sum\limits_{\substack{k=1}}^{K}{\beta_k\boldsymbol{U_l}+\frac{1}{M}\left(\sum_{j=1}^{K}\beta_j\lambda''_{lj}+\frac{1}{\rho_p}\right)\boldsymbol{V_l}}
\end{flalign}
\normalsize
where 
\small
\begin{flalign}
\hat \gamma''_{ljk}(a-n)=\int_{-\infty}^{\infty}\int_{-\infty}^{\infty}{\left(\lambda_{lj}^s \boldsymbol{\hat{G}_{lk}}(a,n)\right)^2f(\tau_j)f(\tau_k)d\tau_jd\tau_k}
\end{flalign}
\begin{flalign}
\lambda''_{lk}=\int_{-\infty}^{\infty}{\left(\lambda_{lk}^s \right)^2f(\tau_k)d\tau_k}
\end{flalign}
\normalsize
where $\boldsymbol{\hat{G}_{lk}}={\boldsymbol{W}_{ll}\boldsymbol{\hat G_{k}}}$ and $\boldsymbol{\hat G_{k}}=[(\boldsymbol{G^1_{k}})^T,\cdots,(\boldsymbol{G^K_{k}})^T]^T$. Assuming
the same distribution for all the time delays, the receive antenna index is discarded. Also  $\boldsymbol{U_l}=\boldsymbol{W_{ll}}E[\boldsymbol{\hat G_{k}}\boldsymbol{\hat G}^H_{\boldsymbol{k}}]\boldsymbol{W}^H_{\boldsymbol{ll}}$ and $\boldsymbol{V_l}=\boldsymbol{W_{ll}}\boldsymbol{\Sigma_{n}}\boldsymbol{W}^H_{\boldsymbol{ll}}$. Inserting these values into Eq. (\ref{far3}), will conclude the proof.

\section{}\label{appendix_approx}
In this section, we drive an upper-bound for the approximation mentioned in Appendix \ref{appendix_main}. In other words, we find an upper-bound on the difference between $R_l(a)$ and $R^{approx}_l(a)$ mentioned in Eqs. (\ref{asli}) and (\ref{asli2}), respectively. To find the upper-bound, we use the following lemma:
\begin{lemma}\label{mohem}
Given any two positive random variables, X and Y, we have the following identity:

\small
\begin{flalign}
\left|E\left[\log_2\left(1+\frac{X}{Y}\right)\right]-\log_2\left(1+\frac{E[X]}{E[Y]}\right)\right|\leq \log_2\left(E[X+Y]E\left[\frac{1}{X+Y}\right]E[Y]E\left[\frac{1}{Y}\right]\right)
\label{lemmaa}
\end{flalign}
\normalsize

\end{lemma}

\begin{IEEEproof}
We know that $f(x)=\log_2(x)$ and $g(x)=\log_2(\frac{1}{x})$ are concave and convex functions, respectively. Hence, by using Jensen's inequality, we can get the following bounds:
\begin{flalign}
\log_2\left(\frac{1}{E\left[\frac{1}{Y}\right]}\right)&\leq E\left[\log_2(Y)\right]\leq \log_2\left(E[Y]\right)\\
\log_2\left(\frac{1}{E\left[\frac{1}{X+Y}\right]}\right)&\leq E\left[\log_2\left(X+Y\right)\right]\leq \log_2\left(E[X+Y]\right)
\end{flalign}
By combining these inequalities, we can conclude that:

\small
\begin{flalign}
\log_2\left(\frac{1}{E\left[\frac{1}{X+Y}\right]}\right)-\log_2\left(E[Y]\right)&\leq E\left[\log_2(X+Y)\right]-E\left[\log_2(Y)\right]\leq \log_2(E[X+Y])-\log_2\left(\frac{1}{E\left[\frac{1}{Y}\right]}\right)\\
\log_2\left(\frac{1}{E\left[\frac{1}{X+Y}\right]}\right)-\log_2(E[Y])&\leq \log_2(E[X+Y])-\log_2(E[Y])\leq \log_2(E[X+Y])-\log_2\left(\frac{1}{E\left[\frac{1}{Y}\right]}\right)
\end{flalign}
\normalsize

We know that if $A\leq x \leq B$ and $A\leq y \leq B$, then $|x-y|\leq B-A$. Therefore, $\left|E\left[\log_2\left(1+\frac{X}{Y}\right)\right]-\log_2\left(1+\frac{E[X]}{E[Y]}\right)\right|$ is upper-bounded by $\log_2\left(E[X+Y]\right)-\log_2\left(\frac{1}{E\left[\frac{1}{Y}\right]}\right)-\log_2\left(\frac{1}{E\left[\frac{1}{X+Y}\right]}\right)+\log_2\left(E[Y]\right)$. After some calculations, the upper-bound can be simplified to $\log_2\left(E[X+Y]E\left[\frac{1}{X+Y}\right]E[Y]E\left[\frac{1}{Y}\right]\right)$ which completes the proof.
\end{IEEEproof}

By applying a Taylor series expansion of $\frac{1}{Y}$ around $E[Y]$,
we will have:
\begin{flalign}\label{roomokh}
E\left[\frac{1}{Y}\right]= \frac{1}{E[Y]}+O\left(\frac{VAR[Y]}{E^3[Y]}\right)
\end{flalign}
as $\frac{VAR[Y]}{E[Y]}\rightarrow 0$\cite{rate_massiveMIMO}. Eq. (\ref{roomokh}), implies that there exist positive numbers $\epsilon$ and $C$ such that:
\begin{flalign}
E\left[\frac{1}{Y}\right]\leq \frac{1}{E[Y]}+C\frac{VAR[Y]}{E^3[Y]} \ \ \ \ when \ \frac{VAR[Y]}{E[Y]}\leq \epsilon
\end{flalign}
Therefore, the inequality in Eq. (\ref{lemmaa}), can be rewritten as:

\small
\begin{flalign}
\left|E\left[\log_2\left(1+\frac{X}{Y}\right)\right]-\log_2\left(1+\frac{E[X]}{E[Y]}\right)\right|\leq \log_2\left( \left(1+C_1\frac{VAR[X+Y]}{E^2[X+Y]}\right)\left(1+C_2\frac{VAR[Y]}{E^2[Y]}\right)\right)
\end{flalign}
\normalsize

In Eq. (\ref{asli}), $\boldsymbol{T_{lk}}(a,n)$ is equal to the average of $M$ independent R.V.s, i.e., $\boldsymbol{T_{lk}}(a,n)=\frac{1}{M}\sum_{i=1}^{M}{t_{lk}^i(a,n)}$. The expected value and variance of each summand is denoted as $\mu_{lk}(a,n)$ and $\sigma^2_{lk}(a,n)$, respectively. As a result, the expected value and variance of $\boldsymbol{T_{lk}}(a,n)$ are equal to $\mu_{lk}(a,n)$ and $\frac{1}{M}\sigma^2_{lk}(a,n)$, respectively. Then, $E[|\boldsymbol{T_{lk}}(a,n)|^2]$ and $VAR[|\boldsymbol{T_{lk}}(a,n)|^2]$ can be bounded by:
\begin{flalign}
\label{tay}
VAR[|\boldsymbol{T_{lk}}(a,n)|^2]&\leq \frac{4}{M}\mu^2_{lk}(a,n)\sigma^2_{lk}(a,n)\\
\label{gen}
E[|\boldsymbol{T_{lk}}(a,n)|^2]&\geq \mu^2_{lk}(a,n)
\end{flalign}
where Inequality (\ref{tay}) is derived using Taylor approximation, i.e., $VAR[f(X)]=(f'(E[X]))^2VAR[X]-\frac{(f''(E[X]))^2}{4}VAR^2[X]$ and Inequality (\ref{gen}) is a result of Jensen's inequality. After some calculations, it can be shown that $\frac{VAR[Y]}{E[Y]}$ and $\frac{VAR[X+Y]}{E[X+Y]}$ can be made sufficiently small by increasing $M$, i.e, the number of receiver antennas, and as a result we have:
\begin{flalign}\label{taghrib}
|R_{l}(a)-R^{approx}_{l}(a)|\leq 2\log_2\left(1+\frac{c\sigma^2}{M\mu^2}\right)
\end{flalign}

where $\mu=\min_{l,k,n}{ \mu^2_{lk}(a,n)}$, $\sigma^2=\max_{l,k,n}{4\mu^2_{lk}(a,n)\sigma^2_{lk}(a,n)}$ and $c$ is a constant. When $M$ grows large, the approximation becomes tighter and $ \tilde{R}_{l}(a)\rightarrow R_{l}(a)$ as $M$ goes to infinity which is line with the fact that variables get close to determinitic values as $M$ grows large. Simulation results also show that for $M$ larger than 100, the approximation is very precise. \\



\bibliographystyle{IEEEtran}
\bibliography{reference}

\begin{thebibliography}{10}
\providecommand{\url}[1]{#1}
\csname url@samestyle\endcsname
\providecommand{\newblock}{\relax}
\providecommand{\bibinfo}[2]{#2}
\providecommand{\BIBentrySTDinterwordspacing}{\spaceskip=0pt\relax}
\providecommand{\BIBentryALTinterwordstretchfactor}{4}
\providecommand{\BIBentryALTinterwordspacing}{\spaceskip=\fontdimen2\font plus
\BIBentryALTinterwordstretchfactor\fontdimen3\font minus
  \fontdimen4\font\relax}
\providecommand{\BIBforeignlanguage}[2]{{%
\expandafter\ifx\csname l@#1\endcsname\relax
\typeout{** WARNING: IEEEtran.bst: No hyphenation pattern has been}%
\typeout{** loaded for the language `#1'. Using the pattern for}%
\typeout{** the default language instead.}%
\else
\language=\csname l@#1\endcsname
\fi
#2}}
\providecommand{\BIBdecl}{\relax}
\BIBdecl

\bibitem{bolcskei2006space}
H.~B{\"o}lcskei, \emph{Space-time wireless systems: from array processing to
  {MIMO} communications}.\hskip 1em plus 0.5em minus 0.4em\relax Cambridge
  University Press, 2006.

\bibitem{duman2008coding}
T.~M. Duman and A.~Ghrayeb, \emph{Coding for {MIMO} communication
  systems}.\hskip 1em plus 0.5em minus 0.4em\relax John Wiley \& Sons, 2008.

\bibitem{jafarkhani2005space}
H.~Jafarkhani, \emph{Space-time coding: theory and practice}.\hskip 1em plus
  0.5em minus 0.4em\relax Cambridge university press, 2005.

\bibitem{foschini1996layered}
G.~J. Foschini, ``Layered space-time architecture for wireless communication in
  a fading environment when using multi-element antennas,'' \emph{Bell Labs
  Technical Journal}, vol.~1, no.~2, pp. 41--59, Autumn 1996.

\bibitem{tarokh1999space}
V.~Tarokh, H.~Jafarkhani, and A.~R. Calderbank, ``Space-time block codes from
  orthogonal designs,'' \emph{IEEE Transactions on Information Theory},
  vol.~45, no.~5, pp. 1456--1467, Jul. 1999.

\bibitem{verdu1998multiuser}
S.~Verdu, \emph{{Multiuser detection}}.\hskip 1em plus 0.5em minus 0.4em\relax
  Cambridge University Press, 1998.

\bibitem{van1999time}
J.~J. van~de Beek, P.~O. Borjesson, M.~L. Boucheret, D.~Landstrom, J.~M.
  Arenas, P.~Odling, C.~Ostberg, M.~Wahlqvist, and S.~K. Wilson, ``A time and
  frequency synchronization scheme for multiuser {OFDM},'' \emph{IEEE Journal
  on Selected Areas in Communications}, vol.~17, no.~11, pp. 1900--1914, Nov.
  1999.

\bibitem{zhou2005synchronization}
E.~Zhou, X.~Zhang, H.~Zhao, and W.~Wang, ``Synchronization algorithms for
  {MIMO} {OFDM} systems,'' in \emph{IEEE Wireless Communications and Networking
  Conference, 2005}, vol.~1, Mar. 2005, pp. 18--22 Vol. 1.

\bibitem{timing_recovery}
A.~A. Nasir, S.~Durrani, and R.~A. Kennedy, ``Blind timing and carrier
  synchronization in distributed multiple input multiple output communication
  systems,'' \emph{IET Communications}, vol.~5, no.~7, pp. 1028--1037, May
  2011.

\bibitem{das2011mimo}
A.~Das and B.~D. Rao, ``{MIMO} systems with intentional timing offset,''
  \emph{EURASIP Journal on Advances in Signal Processing}, vol. 2011, no.~1,
  pp. 1--14, 2011.

\bibitem{lu2012asynchronous}
L.~Lu and S.~C. Liew, ``Asynchronous physical-layer network coding,''
  \emph{IEEE Transactions on Wireless Communications}, vol.~11, no.~2, pp.
  819--831, Feb. 2012.

\bibitem{poorkasmaei2015asynchronous}
S.~Poorkasmaei and H.~Jafarkhani, ``Asynchronous orthogonal differential
  decoding for multiple access channels,'' \emph{IEEE Transactions on Wireless
  Communications}, vol.~14, no.~1, pp. 481--493, Jan. 2015.

\bibitem{avendi2015differential}
M.~R. Avendi and H.~Jafarkhani, ``Differential distributed space-time coding
  with imperfect synchronization in frequency-selective channels,'' \emph{IEEE
  Transactions on Wireless Communications}, vol.~14, no.~4, pp. 1811--1822,
  Apr. 2015.

\bibitem{mehdi}
M.~Ganji and H.~Jafarkhani, ``Interference mitigation using asynchronous
  transmission and sampling diversity,'' in \emph{2016 IEEE Global
  Communications Conference (GLOBECOM)}, Dec. 2016, pp. 1--6.

\bibitem{massiveMIMO}
T.~L. Marzetta, ``Noncooperative cellular wireless with unlimited numbers of
  base station antennas,'' \emph{IEEE Transactions on Wireless Communications},
  vol.~9, no.~11, pp. 3590--3600, Nov. 2010.

\bibitem{larsson2014massive}
E.~G. Larsson, O.~Edfors, F.~Tufvesson, and T.~L. Marzetta, ``Massive {MIMO}
  for next generation wireless systems,'' \emph{IEEE Communications Magazine},
  vol.~52, no.~2, pp. 186--195, Feb. 2014.

\bibitem{ngo2013energy}
H.~Q. Ngo, E.~G. Larsson, and T.~L. Marzetta, ``Energy and spectral efficiency
  of very large multiuser {MIMO} systems,'' \emph{IEEE Transactions on
  Communications}, vol.~61, no.~4, pp. 1436--1449, Apr. 2013.

\bibitem{lu2014overview}
L.~Lu, G.~Y. Li, A.~L. Swindlehurst, A.~Ashikhmin, and R.~Zhang, ``An overview
  of massive {MIMO}: Benefits and challenges,'' \emph{IEEE Journal of Selected
  Topics in Signal Processing}, vol.~8, no.~5, pp. 742--758, Oct. 2014.

\bibitem{rusek2013scaling}
F.~Rusek, D.~Persson, B.~K. Lau, E.~G. Larsson, T.~L. Marzetta, O.~Edfors, and
  F.~Tufvesson, ``Scaling up {MIMO}: Opportunities and challenges with very
  large arrays,'' \emph{IEEE Signal Processing Magazine}, vol.~30, no.~1, pp.
  40--60, Jan. 2013.

\bibitem{rogalin2014scalable}
R.~Rogalin, O.~Y. Bursalioglu, H.~Papadopoulos, G.~Caire, A.~F. Molisch,
  A.~Michaloliakos, V.~Balan, and K.~Psounis, ``Scalable synchronization and
  reciprocity calibration for distributed multiuser {MIMO},'' \emph{IEEE
  Transactions on Wireless Communications}, vol.~13, no.~4, pp. 1815--1831,
  Apr. 2014.

\bibitem{motivation}
H.~Zhang, N.~B. Mehta, A.~F. Molisch, J.~Zhang, and S.~H. Dai, ``Asynchronous
  interference mitigation in cooperative base station systems,'' \emph{IEEE
  Transactions on Wireless Communications}, vol.~7, no.~1, pp. 155--165, Jan
  2008.

\bibitem{pitarokoilis2012effect}
A.~Pitarokoilis, S.~K. Mohammed, and E.~G. Larsson, ``Effect of oscillator
  phase noise on uplink performance of large {MU-MIMO} systems,'' in \emph{2012
  50th Annual Allerton Conference on Communication, Control, and Computing
  (Allerton)}, Oct. 2012, pp. 1190--1197.

\bibitem{krishnan2016linear}
R.~Krishnan, M.~R. Khanzadi, N.~Krishnan, Y.~Wu, A.~G. i~Amat, T.~Eriksson, and
  R.~Schober, ``Linear massive {MIMO} precoders in the presence of phase
  noise-a large-scale analysis,'' \emph{IEEE Transactions on Vehicular
  Technology}, vol.~65, no.~5, pp. 3057--3071, May 2016.

\bibitem{pitarokoilis2012optimality}
A.~Pitarokoilis, S.~K. Mohammed, and E.~G. Larsson, ``On the optimality of
  single-carrier transmission in large-scale antenna systems,'' \emph{IEEE
  Wireless Communications Letters}, vol.~1, no.~4, pp. 276--279, Aug. 2012.

\bibitem{rate_massiveMIMO}
Q.~Zhang, S.~Jin, K.~K. Wong, H.~Zhu, and M.~Matthaiou, ``Power scaling of
  uplink massive {MIMO} systems with arbitrary-rank channel means,'' \emph{IEEE
  Journal of Selected Topics in Signal Processing}, vol.~8, no.~5, pp.
  966--981, Oct. 2014.

\bibitem{hassibi}
B.~Hassibi and B.~M. Hochwald, ``How much training is needed in
  multiple-antenna wireless links?'' \emph{IEEE Transactions on Information
  Theory}, vol.~49, no.~4, pp. 951--963, Apr. 2003.

\end{thebibliography}

%






\end{document}